\documentclass[aps,pra,reprint,amssymb,showpacs,twocolumns,superscriptaddress,notitlepage]{revtex4-2} 
\usepackage{bm}
\usepackage{graphicx}
\usepackage{dcolumn}
\usepackage{braket}
\usepackage{bbold}
\usepackage{natbib}
\usepackage{amsmath}
\usepackage{color}
\usepackage{dsfont}
\usepackage{comment}
\usepackage{ragged2e}
\usepackage{tikz}
\usepackage{adjustbox}

\definecolor{mayablue}{rgb}{0.45, 0.76, 0.98}
\definecolor{deepskyblue}{rgb}{0.0, 0.75, 1.0}
\definecolor{dodgerblue}{rgb}{0.12, 0.56, 1.0}
\definecolor{ultramarineblue}{rgb}{0.25, 0.4, 0.96}
\definecolor{portlandorange}{rgb}{1.0, 0.35, 0.21}
\definecolor{purple(x11)}{rgb}{0.63, 0.36, 0.94}
\definecolor{gold}{rgb}{0.99, 0.76, 0.0}

\definecolor{forestgreen}{RGB}{34,139,34}

\usepackage[colorlinks]{hyperref}
\hypersetup{   
	linkcolor=black, 
	urlcolor=blue,
    citecolor=blue,
	pdfstartview=
}
\usepackage{orcidlink}

\begin{document}

\title{A single-photon microwave switch with recoverable control photon}

\author{Davide Rinaldi\,\orcidlink{0009-0002-2562-0807}} 
\affiliation{Dipartimento di Fisica, Universit\`a di Pavia, via Bassi 6, 27100 Pavia}

\author{Davide Nigro\,\orcidlink{0000-0002-5689-0650}}
\affiliation{Dipartimento di Fisica, Universit\`a di Pavia, via Bassi 6, 27100 Pavia}

\author{Dario Gerace\,\orcidlink{0000-0002-7442-125X}}
\affiliation{Dipartimento di Fisica, Universit\`a di Pavia, via Bassi 6, 27100 Pavia}

\begin{abstract}
{Scalable quantum technologies may be applied in prospective architectures employing traditional information processing elements, such as transistors, rectifiers, or switches modulated by low-power inputs. In this respect, recently developed quantum processors based, e.g., on superconducting circuits may alternatively be employed as the basic platform for ultra-low-power consumption classical processors, in addition to obvious applications in quantum information processing and quantum computing. Here we propose a single-photon microwave switch based on a circuit quantum electrodynamics setup, in which a single control photon in a transmission line is able to switch on/off the propagation of another single photon in a separate line.} {The performances of this single-photon switch are quantified in terms of the photon flux through the output channel, providing a direct comparison of our results with available data. Furthermore, we show how the design of this microwave switch enables the recovery of the single control photon after the switching process.}  This proposal may be readily realized in state-of-art superconducting circuit technology.
\end{abstract}

\maketitle

\section{Introduction}
Several emerging quantum technologies allow to achieve a high degree of control within networks consisting of local quantum systems coupled through propagating channels, also referred to as quantum input-output networks. In this context, we hereby propose a single-photon switching device based on the implementation of the Jaynes-Cummings (JC) model in an open quantum optical circuit. We envision its realization in a superconducting platform, exploiting state-of-art technology that is currently developed to realize quantum processors. In fact, we hereby enforce the idea that these systems may be employed for prospective applications even in the classical information realm with ultra-low-power consumption requirements, such as triggering a single microwave photon signal through a single photon control. \\
In order to take into account both the temporal dependence of the single-photon wave packets and the open nature of the quantum systems within an input/output formalism, the theoretical analysis is carried out by means of the so-called SLH framework \cite{Combes_SLH_framework}, which is a theoretical framework efficiently combining the input-output theory approach \cite{Gardiner:1993, GardinerZoller, Charmichael} with a network-like description \cite{GoughJames_2008, Gough_James_2009}. Such an approach is particularly useful when modelling, as in our case, classical information processing networks employing quantum signals, for which photon shot noise cannot be ignored due to the low-excitation content of each signal. These networks become more and more relevant as electronic devices for classical information processing become prone to heat dissipation and power consumption, or in other applications related to the communication and sensing domain specifically tailored for the microwave regime \cite{CasariegoQST2023}. \\
Along these lines, previous works have analysed various configurations in which single or coupled quantum systems are employed as single-photon transistors or switches \cite{hartmann2013single,manzoni2014,YanPRA2014,xu2016single,Kyriienko2016,StolyarovPRA2020, wilsmann2018control}. Despite several years of theoretical proposals, a single-photon transistor in the microwave regime has only been recently realized with good performances, by exploiting two superconducting resonators coupled to a single qubit \cite{Wang2022}. In the latter work, a classical signal pulse containing an average of 37 photons is successfully controlled by a single microwave photon. At difference with these early attempts, here we address the switching of a single-photon pulse by a single control photon, i.e., the ultimate regime of ultra-low power classical information processing. Our scheme of operation is based on an idea originally implemented in the optical domain \cite{volz2012ultrafast}, to which we add the unique possibilities offered by superconducting circuits: we assume the signal single-photon pulse to be coupled to the harmonic oscillator, while the control single-photon is directly coupled to the qubit.  In addition, we propose an effective protocol to recover the control photon for later re-usage, thus minimizing the loss of information and power consumption. 
Despite its simplicity, our proposed protocol allows to achieve an optimal switching figure of merit, which is promising in view of experimental realizations in state-of-art superconducting quantum circuits. In fact, the switching performances of this device ultimately depend on the cavity-qubit coupling rate as compared to the loss rates; the unpaired figures of merit already reached in superconducting circuits in this respect make the latter (and, as a consequence, the microwave domain) the preferential choice for the actual implementation of our proposed scheme.

\section{Single-photon switching scheme}\label{sec:switch_description}

\begin{figure}[t]
\centering
    \includegraphics[width=0.5\textwidth]{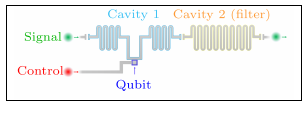}
    \caption{Scheme of the input-output network for single-photon switching with a single control photon.}
    \label{fig: fig1B}
\end{figure}

\begin{figure}[t]
    \includegraphics[width=0.5\textwidth]{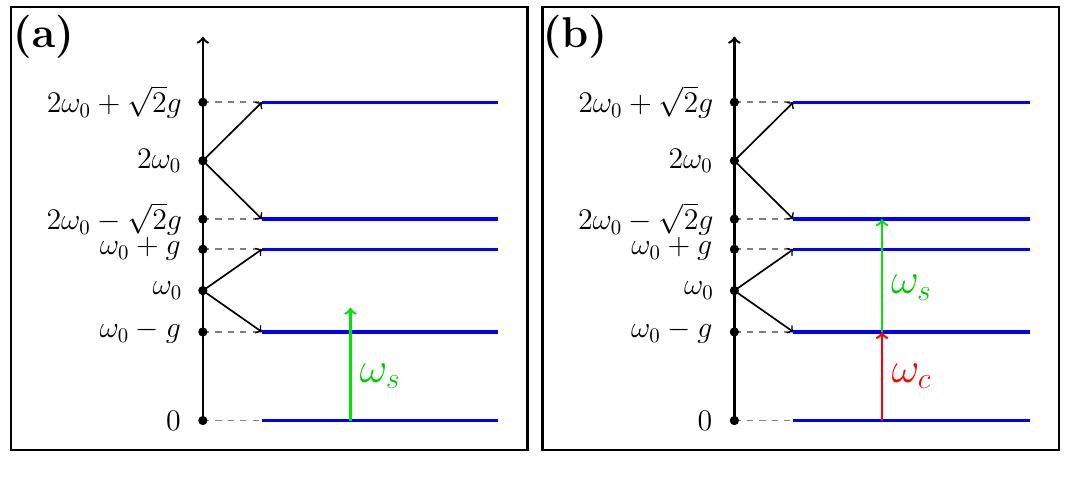}
    \caption{Schematic representation of the JC ladder spectrum, with the choice of single-photon signal ($\omega_s$) and control ($\omega_c$) frequencies employed in this work to achieve an efficient single-photon switching.}
    \label{fig: fig1C}
\end{figure}

In this Section we briefly describe the principle of operation of our single-photon microwave switching device. The basic structure is schematically depicted in Fig.~\ref{fig: fig1B}, consisting of a pair of superconducting transmission line resonators and a superconducting qubit, denoted as Cavity 1, Cavity 2, and Qubit, respectively. As a starting point, we begin by describing the Cavity 1 + Qubit parts of the network. When Cavity 1 is strongly coupled to the Qubit, the absorption of a single \textit{signal} photon with carrier frequency $\omega_s$ into Cavity 1, and its subsequent transmission into the output channel, can be controlled by shining a second single-photon pulse (Control) with frequency $\omega_c$ onto the Qubit. This behavior results from the peculiar energy spectrum of the Cavity 1 + Qubit subsystem. Let $g$ denote the coherent coupling rate between Cavity 1 and  Qubit, if $\omega_{\text{cavity}}$ and $\omega_{\text{qubit}}$ arethe bare transition frequencies of the two oscillators, respectively, the spectrum of excitations above their ground state level is accurately described by the Jaynes-Cummings (JC) ladder \cite{Scully_Zubairy_book1997}. Suitably engineered superconducting circuits have  been experimentally shown to have this spectral solution \cite{schuster2007resolving,CircuitQED}. In fact, the JC Hamiltonian in rotating wave approximation can be analytically diagonalized, leading to an expression for the eigenenergies of the $N$ excitations eigenstates, which reads
\begin{equation}
    \label{eq: JC ladder}
\begin{split}
    \hbar\omega_{\pm}(N) &= (N-1)\hbar\omega_{\text{cavity}} + \frac{\hbar\omega_{\text{cavity}}+\hbar\omega_{\text{qubit}}}{2} \\
    &\pm \frac{1}{2} \sqrt{\hbar^2\Delta^2 + 4\hbar^2g^2N} \, ,
\end{split}
\end{equation}
with  $\Delta = \omega_{\text{cavity}} - \omega_{\text{qubit}}$. Given the type of input states considered in the present work, the system response is fully understood by looking at the JC levels belonging to the $N\leq 2$ subset. A sketch of such levels for $\omega_{\text{cavity}} = \omega_{\text{qubit}} \equiv \omega_0$ (which is the case addressed in the present analysis) is reported in the two panels of Fig.~\ref{fig: fig1C}. Due to the frequency mismatch between $\omega_s = \omega_0 - (\sqrt{2}-1)g$ and the eigenenergies of the $N=1$ subset, as shown in  Fig.~\ref{fig: fig1C}(a), the signal photon alone cannot induce any single-photon transition in the system. Therefore, no signal emerges from the output of the structure if only a single signal photon drives Cavity 1  (independently of the presence of Cavity 2); the overall network then behaves like an open switch, i.e., corresponding to an OFF state. On the other hand, if the Cavity 1 + Qubit system is also excited with a control photon at frequency $\omega_c = \omega_0 - g$, then the signal photon is resonant with the transition from $\hbar \omega_-(N=1)$ to $\hbar \omega_-(N=2)$, and it can be absorbed by Cavity 1, as sketched in Fig.~ \ref{fig: fig1C}(b). 

{By suitably filtering the emission from Cavity 1, e.g., by directly coupling it with Cavity 2, the signal photon can be collected at the output of the network. In the latter case, the single-photon switch is in the ON state. In addition, we notice that a third possible situation exists in this setup, in which the control photon alone is absorbed within the Cavity 1 + Qubit subsystem. In this respect, the presence of the filtering Cavity 2 is able to suppress the transmission of such photons, due to the mismatch between $\omega_c$  and the characteristic frequency of such a filtering component.
The overall switching behavior, which we are going to theoretically describe in detail in the following, is summarized in Fig. 3.}

\section{Theoretical model}
\begin{figure}[t]
    \includegraphics[width=0.5\textwidth]{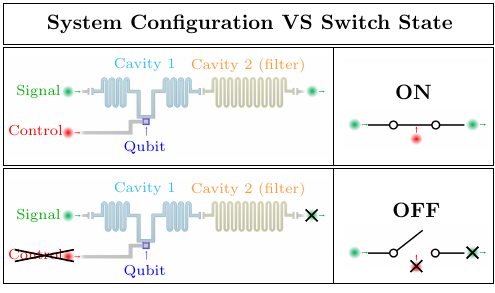}
    \caption{Schematic illustration of the single-photon microwave switching circuit implemented in a cavity quantum electrodynamics setup, with the description of the ON (top panels) and OFF (bottom panels) working configurations. }
    \label{fig: fig1A}
\end{figure}

Here we summarize the main ingredients used in the theoretical characterization of the setup sketched in Fig.~\ref{fig: fig1A}. We first define the relevant parameters and the Hamiltonian model used in the following. Then, we introduce the SLH framework, used in the present work to account for the coupling of the localized network to the input and output channels. The input-output network used to model the two configurations (without and with filtering cavity, respectively) is illustrated in Fig.~\ref{fig:model}, and further details are reported in Apps.~\ref{appendix: theory in details},  \ref{appendix: slh framework} and \ref{appendix: slh equations}. \\

\subsection{The system Hamiltonian}
\begin{figure}[t]
    \centering
    \includegraphics[width=0.5\textwidth]{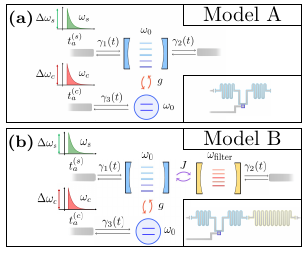}
    \caption{(a) Input-output network describing Cavity 1 (light blue, with harmonic energy spectrum) coupled to the Qubit (blue, two-level system), with the relevant simulation parameters. It constitutes the core of the device, since its behavior can be described by the JC model, which is the basis of the switching protocol. (b) Extension of the model by the addition of Cavity 2 (yellow), which acts as a filter for the control photon.}
    \label{fig:model}
\end{figure}

In this Section we summarize the theoretical parametrization of the systems constituting the building blocks of the quantum network. Details concerning the input-output channels, their coupling to the network, as well as the SLH framework are provided in the following. We will focus on the response of two different system configurations, as illustrated in Fig.~\ref{fig:model}. The first configuration, corresponding to ``Model A" in Fig.~\ref{fig:model}(a), accounts for a simplified network consisting of Cavity 1 and the qubit, with the respective input-output channels. The second one, "Model B" (see Fig.~\ref{fig:model}(b)), includes an extra cavity in the network, i.e. Cavity 2, playing the role of a filter used to select the right frequency contribution at the output of the network.

By means of a standard second quantization formalism, the total Hamiltonian describing ``Model A" can be expressed as
\begin{equation}
    \label{eq: Hamiltonian-A}
    \hat{H}_{A} = \hat{H}_{\text{free}}^{(\text{cavity 1})} + \hat{H}_{\text{free}}^{(\text{qubit})} +\hat{H}_{\text{int}}^{(\text{cavity 1, qubit})} \, ,
\end{equation}
in which $\hat{H}_{\text{free}}^{(\text{cavity 1})} = \hbar \omega_0 \hat{a}_1^\dagger \hat{a}_1$ and $\hat{H}_{\text{free}}^{(\text{qubit})} = \hbar \omega_0 \hat{\sigma}_+ \hat{\sigma}_-$ represent the bare Cavity 1 and qubit Hamiltonian operators, respectively, while $\hat{H}_{\text{int}}^{(\text{cavity 1, qubit})} = \hbar g (\hat{a}_1^\dagger\hat{\sigma}_- + \hat{a}_1\hat{\sigma}_+)$ represents their mutual coupling in the rotating wave approximation, quantified by the interaction strength, $g$.
Given $\hat{H}_A$, the Hamiltonian for ``Model B" can be straightforwardly formulated as
\begin{equation}
    \label{eq: Hamiltonian-B}
    \hat{H}_{B}=\hat{H}_A+\hat{H}_{\text{free}}^{(\text{cavity 2})} +\hat{H}_{\text{int}}^{(\text{cavity 1, cavity 2})} \, ,
\end{equation}
in which $\hat{H}_{\text{free}}^{(\text{cavity 2})}= \hbar \omega_s \hat{a}_2^\dagger \hat{a}_2$ and $\hat{H}_{\text{int}}^{(\text{cavity 1, cavity2})} = \hbar J (\hat{a}_1^\dagger \hat{a}_2 + \hat{a}_1 \hat{a}_2^\dagger)$, and $J$ is a (weak) coupling constant between the two resonators (e.g., due to evanescent field overlap).

\subsection{Input-output channels and system-bath coupling}
In agreement with standard input-output theory \cite{GardinerCollett, Gardiner:1993}, waveguides used to feed in and extract photons from the localized system are described as a continuum of bosonic degrees of freedom. In the present case, where the signal (s) and control (c) photons are assumed to propagate into two different input channels, and radiation is collected via an output (o) waveguide, the Hamiltonian terms accounting for the field degrees of freedom can be expressed as

\begin{equation}
    \label{eq: Bath Hamiltonian}
    \hat{H}_{\text{env}} = \hat{H}^{(s)}_{\text{free}} + \hat{H}^{(c)}_{\text{free}} +\hat{H}^{(o)}_{\text{free}} + \hat{H}^{\text{sys-env}}_{\text{X}} \, ,
\end{equation}
in which the terms $\hat{H}^{(\alpha)}_{\text{free}} = \int_0^{+\infty} \hbar\omega \hat{b}^{\dagger}_{\alpha}(\omega) \hat{b}_{\alpha}(\omega) d\omega $ describe the free evolution of photons occupying the input and output waveguide eigenmodes. In particular, $\hat{b}_p(\omega)$ ($\hat{b}^{\dagger}_p(\omega)$) are the field annihilation (creation) operators in the frequency domain, with $\alpha = s,c, o$. In this respect, we refer to the field as the environment, which is dispersively coupled to the system of interest (i.e. the switch itself). In turn, the system of interest is composed of multiple subsystems (i.e., Cavity 1, Cavity 2 and the qubit), that are called localized systems. The interaction part $\hat{H}^{\text{sys-env}}_{\text{X}}$ ($\text{X}=\text{A,B}$) represents thus the coupling between the environment and the localized systems. For Model A, such interaction term reads
\begin{equation}
    \label{eq: H bath-sys A}
    \hat{H}^{\text{sys-env}}_{\text{A}} = \hat{H}^{(\text{cavity 1, } s)} +  \hat{H}^{(\text{cavity 1, } o)}_{\text{int}} + \hat{H}^{(\text{qubit, } c)}_{\text{int}} \, ,
\end{equation}
while for Model B we assumed the following 
\begin{equation}
    \label{eq: H bath-sys B}
    \hat{H}^{\text{sys-env}}_{\text{B}} = \hat{H}^{(\text{cavity 1, } s)}  + \hat{H}^{(\text{cavity 2, } o)}_{\text{int}} + \hat{H}^{(\text{qubit, } c)}_{\text{int}} \, .
\end{equation}
In particular, as shown in App.~\ref{appendix: theory in details}, interaction terms are assumed to be linear in both system and environment operators. In this case, if the interaction strength between the input-output channels and the localized system can be assumed to be constant over a frequency range around the signal ($\omega_s$) and control ($\omega_c$) frequencies (see App.~\ref{appendix: theory in details} for further details), the system-environment interaction terms in Eqs.~\ref{eq: H bath-sys A} and \ref{eq: H bath-sys B} reduce to the following expressions
\begin{equation}
    \label{eq: bath-sys interaction Hamiltonians}
    \begin{split}
    &\hat{H}^{(\text{cavity 1, } s)}_{\text{int}} = i\hbar\sqrt{\gamma_1}[\hat{a}_1\hat{b}_s^\dagger(t) - \hat{a}_1^\dagger \hat{b}_s(t)] \\
    &\hat{H}^{(\text{cavity 1, } o)}_{\text{int}} = i\hbar\sqrt{\gamma_2}[\hat{a}_1\hat{b}_o^\dagger(t) - \hat{a}_1^\dagger \hat{b}_o(t)] \\
  &\hat{H}^{(\text{cavity 2, } o)}_{\text{int}} = i\hbar\sqrt{\gamma_2}[\hat{a}_2\hat{b}_o^\dagger(t) - \hat{a}_2^\dagger \hat{b}_o(t)] \\
&\hat{H}^{(\text{qubit, } c)}_{\text{int}} = i\hbar\sqrt{\gamma_3}[\hat{\sigma}_-\hat{b}_c^\dagger(t) - \hat{\sigma}_+ \hat{b}_c(t)] \, ,
    \end{split}
\end{equation}
in which $\hat{b}_p(t)$ ($\hat{b}^{\dagger}_p(t)$) are field operators in time domain ($p=s,c,o$). We notice that with such choice of parameters, independently of the particular model used for the system description, there is a one-to-one correspondence between input-output channels and the effective coupling provided by the rates $\gamma_{i}$.

\subsection{Time evolution in the SLH framework}
In this section we summarize the core ideas at the basis of the time-evolution method employed in this work to analyze the response of the quantum networks described in previous sections. Further details are given in Appendices \ref{appendix: slh framework} and \ref{appendix: slh equations}.
When describing electrical circuits, the network response results from the way the many components are wired together, i.e., concatenated between each other. Similarly, when using the SLH framework  the propagation of signals from the input channels to the output ones is encoded into a triple of operators, defined as ($\hat{\textbf{S}}$, $\hat{\textbf{L}}$, $\hat{H}$), whose elements are determined by the way each localized system operator $X_l$ (in our case, corresponding to Cavity 1, Qubit, or Cavity 2) is connected to its neighbors and/or to the input-output channels. In particular, once the set of triples $\{$($\hat{S}_l$, $\hat{L}_l$, $\hat{H}_l$)$\}$ associated to set of localized system operators $\{X_l\}$ and the network ``topology" are known, the triple of the whole network 
is uniquely determined.

In the present analysis, since no feedback loops are present and neighboring localized subsystems interact via Hamiltonian terms, the description gets simplified. More precisely, the time-evolution is determined by the (rotated) global Hamiltonian ($\hat{H}$) and the coupling operators with the input-output channels ($\hat{\textbf{L}}$). In particular, due to the form of system-environment interaction terms, the components of the vector $\hat{\textbf{L}}$ are always proportional to the creation/annihilation (raising/lowering) operators when considering bosonic (spin) degrees of freedom, respectively. In addition, it is possible to account for the explicit form of single photon pulses injected into the different channels. In the case of a single photon with spectral density function $\xi(t)$, as discussed in Ref.~\cite{Baragiola:2014}, by defining the single-photon state as $\ket{1_{\xi}} = \int dt \xi(t) \hat{b}^\dagger(t)\ket{0}$ at the input, the time-evolution of the state of localized systems in the network, $\hat{\rho}_{1,1}(t)$, can be obtained by solving the following system of differential equations 
\begin{equation}
\begin{aligned}
        \label{eq: Master equation for field in Fock state, example}
        \frac{d}{dt}&\hat{\rho}_{m,n}(t) = -\frac{i}{\hbar}\big[\hat{H},\hat{\rho}_{m,n}(t)\big] +\mathcal{L}[\hat{L}]\hat{\rho}_{m,n}(t) \\
        &+ \sqrt{m}\xi(t)\big[\hat{\rho}_{m-1,n}(t), \hat{L}^{\dagger}\big] \\
        &+\sqrt{n}\xi^{\ast}(t)\big[\hat{L}, \hat{\rho}_{m,n-1}(t) \big] \, ,
\end{aligned}
\end{equation}
in which $\{\hat{\rho}_{m,n}\}$ ($m,n=0,1$) describes a set of 4 density operators, and $\mathcal{L}[\hat{L}]\hat{\rho}$ indicates the Lindbladian superoperator, defined as $\mathcal{L}[\hat{L}]\hat{\rho} \equiv \hat{L}\hat{\rho}\hat{L}^\dagger -\frac{1}{2}(\hat{L}^\dagger\hat{L} \hat{\rho} + \hat{\rho} \hat{L}^\dagger\hat{L})$.

When more than one single-photon inputs are considered (as we will require in the following), the system equations of motion involve a larger number of terms. This is related to the dependence of the state density operator on the number of photons injected in each channel (see, e.g., Appendix \ref{appendix: slh equations} for  explicit expressions). In the present analysis, we consider a network with 3 input-output channels, in which the $\ket{1_{\xi_s}}$ state (single-photon signal pulse) is injected in channel 1, the vacuum $\ket{0}$ in channel 2, and $\ket{1_{\xi_c}}$ (the single-photon control pulse) in channel 3. In this scenario, the state configuration $\hat{\rho}_{101,101}(t)$ evolves according to the equation
\begin{equation}
\begin{aligned}
        \label{eq: Master equation specific case}
        \frac{d}{dt}&\hat{\rho}_{101,101}(t) = -\frac{i}{\hbar}\big[\hat{H},\hat{\rho}_{101,101}(t)\big] +\sum_{j}\mathcal{L}[\hat{L}_j]\hat{\rho}_{101,101}(t) \\
        &+\xi_s(t)\big[\hat{\rho}_{001,101}(t), \hat{L}^{\dagger}_1\big]+\xi_c(t)\big[\hat{\rho}_{100,101}(t), \hat{L}^{\dagger}_3\big] \\
        &+\xi^{\ast}_s(t)\big[\hat{L}_1, \hat{\rho}_{101,001}(t) \big]+\xi^{\ast}_c(t)\big[\hat{L}_3, \hat{\rho}_{101,100}(t) \big] \, ,
\end{aligned}
\end{equation}
where $\hat{L}_1=\sqrt{\gamma_1} \hat{a}_1$, $\hat{L}_3=\sqrt{\gamma_3} \hat{\sigma}_-$, and $\hat{L}_2=\sqrt{\gamma_2} \hat{a}_1$ (Model A) or $\hat{L}_2=\sqrt{\gamma_2} \hat{a}_2$ (Model B), respectively. These differential equations can be numerically solved, starting from the one associated with the lowest labels (i.e., $\hat{\rho}_{000,000}$) and climbing up to $\hat{\rho}_{101,101}$. Once the set of density operators is known, we can compute the expectation values of any system operator $\hat{X}(t)$ at time $t$, i.e.
\begin{equation}
\label{eq: Average of X}
    \langle \hat{X}(t) \rangle \equiv \text{Tr}\bigl[\hat{\rho}_{101,101}(t)\hat{X}\bigr] \, ,
\end{equation}
where $\hat{X}$ is, e.g., the number operator $\hat{a}_i^\dagger\hat{a}_i$, or the projector $\ket{N}\bra{N}$ on the cavity Fock number state $\ket{N}$. 

Within the SLH formalism it is also possible to compute the output average photon flux through a specific channel \cite{Baragiola_article}. In this paper, we will define $\phi_i(t)$ the average photon flux through channel $i$: it can be calculated by employing a set of equations similar to Eqs.~(\ref{eq: Master equation for field in Fock state, example}), whose rigorous definition is reported in Eq.~(\ref{eq: mean photon flux MULTIMODE}). Importantly, we define the integrated photon flux at time $t$ as: 
\begin{equation}
\label{eq: Integrated photon flux}
    \Phi_i(t) \equiv \int_{t_0}^{t}\phi_i(t')dt' \, ,
\end{equation}
in which $t_0$ is the starting integration time. This quantity is particularly relevant when considering time-dependent input pulses, and/or model parameters.

\section{Numerical results and device optimization}

In this Section we are going to summarize our numerical results on (i) the optimization of the input-output network described by ``Model A'', and (ii) a further optimization of the same model and of ``Model B'', performed by introducing \textit{time-dependent} coupling rates $\gamma_i(t)$. The aim of (i) is to achieve the best figures of merit for single-photon blocking and unlocking by a single control photon, in the case of time-independent coupling rates $\gamma_i$. In this context, we focus on ``Model A'' since the goal is to optimize the fundamental photon reflection/transmission effects due to the JC interaction, without any additional component other than Cavity 1 and qubit. The introduction of the time-dependent parameters in (ii) is motivated by the fact that the device performances can significantly increase by properly adjusting each $\gamma_i(t)$ during the injection process. Once that the efficiency of the process is sufficiently enhanced, we then take into account ``Model B'', which involves Cavity 2. The latter is weakly coupled to Cavity 1, in order to avoid a substantial change in the JC energy spectrum, which would prevent the switching operation previously optimized. The additional cavity acts then as a filter, and causes the photon reflection/transmission to behave as a true single-photon switching process. 

In the following, we will describe single-photon pulses in the form of \textit{time reversed} wavepackets $\xi_p(t)$ \cite{wang_scarani}, expressed as
\begin{equation}
    \label{eq: time-reversed wavepacket}
    \xi_p(t) = \begin{cases}
			\sqrt{\Delta\omega_p} e^{\frac{\Delta\omega_p}{2}(t-t_a^{(p)})} & \text{if $t < t_a^{(p)}$}\\
            0 & \text{if $t \geq t_a^{(p)}$} \, ,
		 \end{cases}
\end{equation}
in which $\Delta\omega_p$ and $t_a^{(p)}$ denote the bandwidth and the arrival time of the photon on the input channel $p$. Notice that single microwave photons can be produced with an almost arbitrary pulse shape  \cite{forn2017demand}, by adding to the device further components such as other superconducting qubits, transmission lines and superconducting quantum interference devices (SQUIDs \cite{granata2016nano}). 
This specific choice of the temporal wavepacket dependence is motivated in App.~\ref{appendix: theory in details}. 

\subsection{Time-independent coupling rates}
As briefly outlined in Sec.~\ref{sec:switch_description}, our proposal for the implementation of an efficient single-photon switch relies on having two photons simultaneously present within the network. Therefore, determining the optimal conditions for which two-particle occupation in the system is maximised represents the first step towards our final goal. Hence, we introduce a proper figure of merit to quantify the system performances. Here we focus on model A, while model B will be discussed in the next subsection. 

\begin{figure}[h!]
    \centering
 \includegraphics[width=0.45\textwidth]{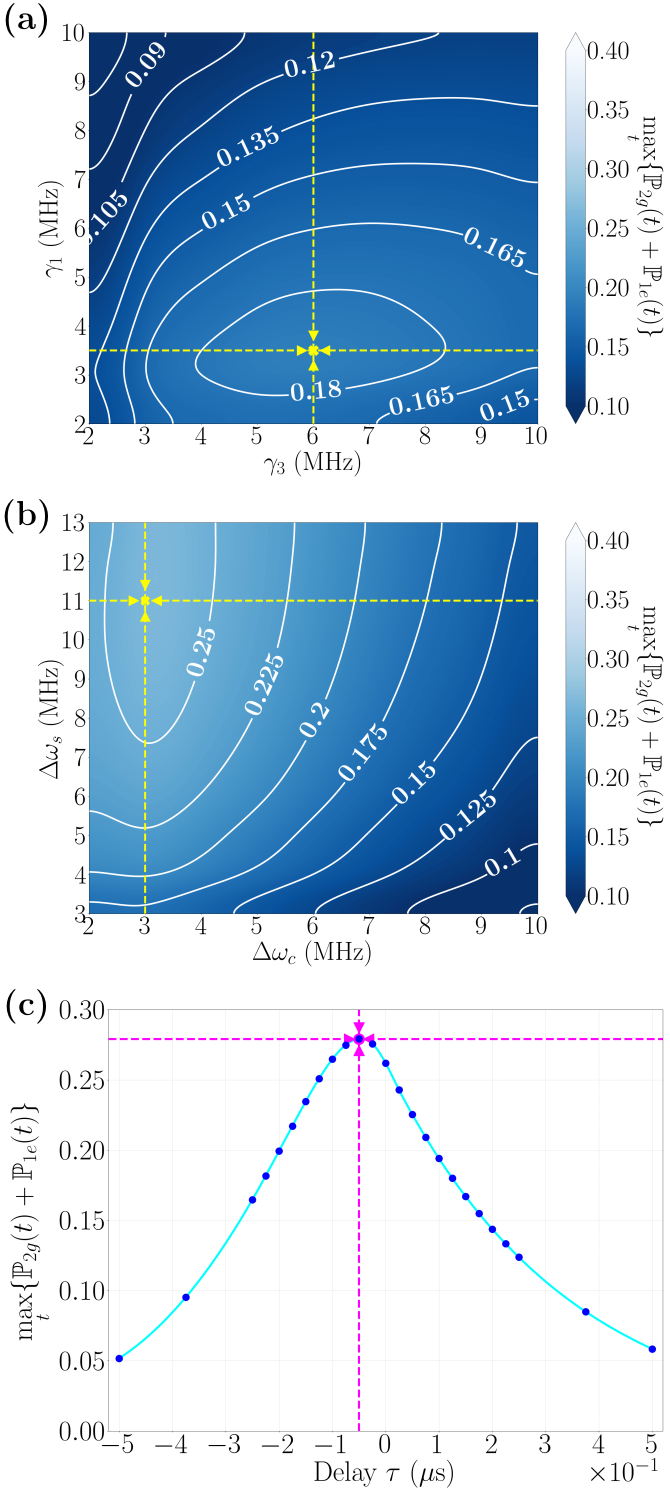}
    \caption{Optimization of the device parameters with respect to the double excitation probability, see Eq.~(\ref{eq: P2}). Each plot shows the maximum value of $\mathds{P}_2(t)$ over the integration time, for a given choice of the model parameters. (a) Optimization of the coupling rates $\gamma_1$ and $\gamma_3$. (b) Optimization of the signal and control single-photon pulse bandwidths,   $\Delta\omega_s$ and $\Delta\omega_c$. (c) Optimization of the delay between the two input pulses, $\tau$. Vertical and horizontal dashed lines are used to help visualizing the location of the best value obtained for the two-particle excitation probability.}
    \label{fig: fig3}
\end{figure}

We consider a further simplified version of the quantum networks introduced in the previous sections, which is obtained by setting $\gamma_2=0$ in model A.  
We are therefore left with two input channels and a reduced system formed by Cavity 1 strongly interacting with the qubit. In this context, the probability of having two excitations in the system at time $t$, $\mathds{P}_2(t)$, is the result of the probability of having a single excitation in the cavity mode and simultaneously the qubit in its excited state, $\mathds{P}_{\text{1e}}(t)$, in addition to the probability of having both excitations stored within the cavity mode while the qubit is in its ground state, $\mathds{P}_{\text{2g}}(t)$. Given the density operator $\hat{\rho}_{101,101}(t)$ at time $t$, it is straightforward to get the overall double excitation probability as
\begin{equation}
    \label{eq: P2}
    \begin{split}
    &\mathds{P}_2(t) = \mathds{P}_{\text{1e}}(t) + \mathds{P}_{\text{2g}}(t) \\ 
    &=  \text{Tr}\bigl[\hat{\rho}_{101,101}(t)\ket{1,e}\bra{1,e}\bigr] + \text{Tr}\bigl[\hat{\rho}_{101,101}(t)\ket{2,g}\bra{2,g}\bigr].   
    \end{split}
\end{equation}

Here is a brief summary of the optimization protocol described in Fig.~\ref{fig: fig3}, and the resulting numerical outcomes. We assumed an initial set of parameters chosen to be as much realistic as possible with respect to typical superconducting quantum circuits platforms \cite{schmidt2013circuit}: we fix $\omega_0 = 4$ GHz and $g = 400$ MHz, initializing the carrier bandwidths such that $\Delta\omega_s = \Delta\omega_c = 4$ MHz, the arrival times at $t_a^{(s)} = t_a^{(c)} = 1.25$~$\mu$s, and $\gamma_1=\gamma_3=6$ MHz. We first explored the parameter space associated to the coupling coefficients $\gamma_1$ and $\gamma_3$. Results concerning the maximal excitation probability $\max\limits_t\{\mathds{P}_2(t)\}$ are shown in Fig.~\ref{fig: fig3}(a). Then, we optimize the system response by varying either the bandwidth or the arrival time of the input photon pulses. Once the optimal $\Delta\omega_s$ and $\Delta\omega_c$ have been found, as shown in Fig.~\ref{fig: fig3}(b), we considered the two-photon excitation probability as a function of the time delay between the arrival times of the two single-photon pulses: $\tau = t_a^{(c)} - t_a^{(s)}$, which is shown in Fig.~\ref{fig: fig3}(c). Finally, the optimization protocol is repeated, this time by initializing the starting parameters equal to the ones provided by the first round. Even after this further optimization, the best results correspond to this very choice of parameters, showing that these represent the optimal choice of parameters. In summary, the optimal result in terms of double-excitation probability is obtained for $\gamma_1 = 3.5$ MHz, $\gamma_3 = 6$ MHz, $\Delta\omega_s = 11$ MHz, $\Delta\omega_c = 3$ MHz, and $\tau = t_a^{(c)} - t_a^{(s)} = -5\cdot10^{-2}$~$\mu$s, which give \textbf{$\max\limits_t\{\mathds{P}_2(t)\} \simeq 0.279$}. Despite the optimization procedure,  $\max\limits_t\{\mathds{P}_2(t)\}$ actually remains relatively low. This is justified by the presence of two loss channels, and two independent driving channels for the control and signal photons. These results are also consistent to what has been already reported, even in the simpler case of a driven cavity-qubit system with single loss channel~\cite{Baragiola:2014}.

After this preliminary optimization, we now explicitly consider in the time evolution the cavity coupling to the output channel by setting $\gamma_2 \neq 0$. Let us define the photon flux integrated up to steady state (see Eq.~\ref{eq: Integrated photon flux}) in the
presence of both signal and control photons, $\Phi_2^{(s,c)}$, and the one
with the control photon only, $\Phi_2^{(c)}$.  We shall also introduce here the steady state integrated photon flux caused in the presence of the signal photon only, $\Phi_2^{(s)}$, which will be useful later. Thus, we are now able to introduce the proper figure of merit quantifying the switching behavior, which is given by the photon flux through channel 2 that is only determined by the signal photon. In practice, we optimized the parameter $\gamma_2$ with respect to the long-time difference between the quantities $\Phi_2^{(s,c)}$ and $\Phi_2^{(c)}$. We assume that this figure of merit provides a quantitative estimate of the contribution at the output of channel 2 that is due to the signal photon only, since any additional flux caused by the passage of the control photon is removed. Hence, we have numerically calculated  these two quantities by simulating the same network first in the presence of both signal and control single-photon wave packets, and then with the control photon only.

\begin{figure}[t]
    \centering
    \includegraphics[width=0.5\textwidth]{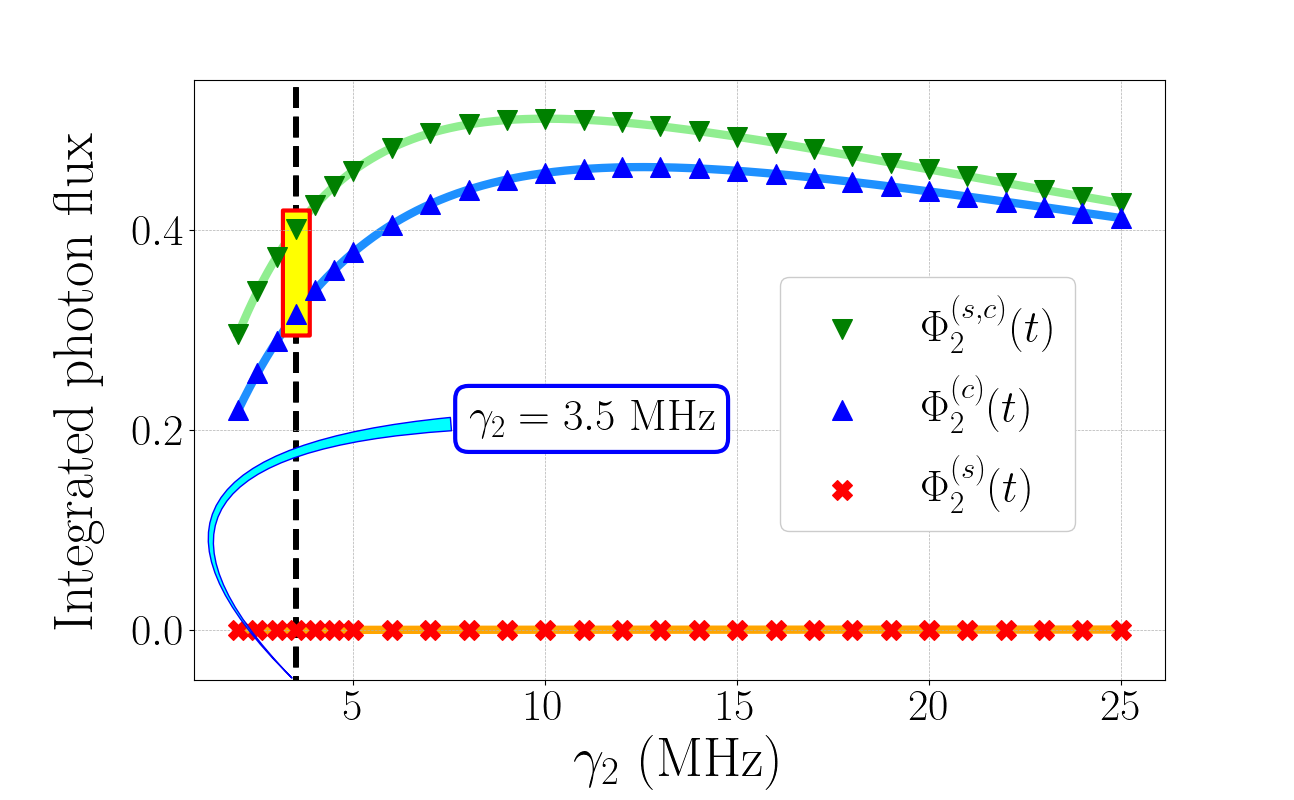}
    \caption{Optimization of $\gamma_2$ with respect to the flux difference at the network output (channel 2), $\Phi_2^{(s,c)} - \Phi_2^{(c)}$, assuming model A. The two highlighted points correspond to the case in which the flux difference is maximized, occurring for $\gamma_2 = 3.5$~MHz.}
    \label{fig: time-INdependent gamma2 opt}
\end{figure}

The optimization of $\gamma_2$ led to the maximum flux difference $\Phi_2^{(s,c)} - \Phi_2^{(c)} = 0.085$ in steady state, corresponding to the optimal value $\gamma_2 = 3.5$ MHz, as depicted in Fig.~\ref{fig: time-INdependent gamma2 opt}.

\subsection{Time-dependent coupling rates}

To further enhance the device functionalities, we then allow for \textit{time-dependent} coupling coefficients, $\gamma_i(t)$. In practice, this could be implemented by using a tunable coupler containing SQUID loops \cite{yin_catch_2013}, which is assumed to be placed between the cavity (or the qubit) and the associated input/output transmission lines. The time evolution of each coupling parameter has been modelled via a simple function, such as a step or square function, in order to emulate a fast switching from $\gamma_i = 0$ to $\gamma_i > 0$, and viceversa, as described by Eqs.~(\ref{eq: time-dep gammas, MAIN}) below. 
As a further working hypothesis, we assume that the optimal parameters $\Delta\omega_s$ and $\Delta\omega_c$ remain unaltered from the time-independent case, which allows to better compare the performances with the results of the previous subsection. Again, we initially restrict our analysis to model A. After having found the optimal parameters for model A, we will change setup and consider a network implementation of model B, thus showing that the presence of a filtering cavity is crucial for the realization of an efficient and realistic single-photon switch. 

\begin{figure}[t]
    \centering
\includegraphics[width=0.45\textwidth]{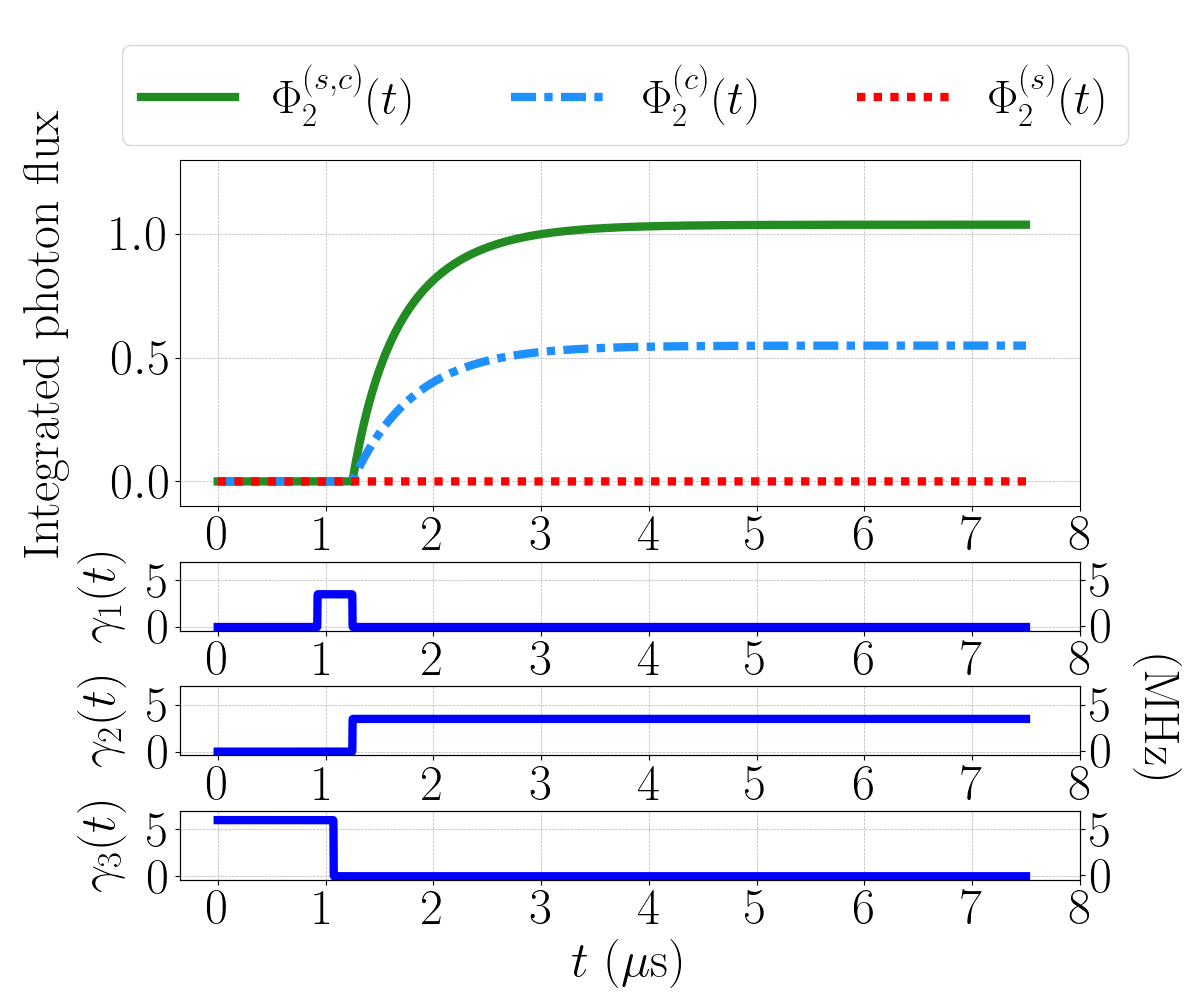}
    \caption{The photon flux through the output of the network in model A, integrated up to time $t$ (see Eq.~\ref{eq: Integrated photon flux}), as a function of the evolution time (i.e., as a function of $t$, corresponding to the integration upper boundary), comparing the following contributions: both signal and control pulses are present, $\Phi_2^{(s,c)}$, only of the control pulse is present, $\Phi_2^{(c)}$, and only the signal is present, $\Phi_2^{(s)}$. In the lower panels, we explicitly report the time-dependent functions $\gamma_i(t)$ as defined in Eq.~(\ref{eq: time-dep gammas, MAIN}), for completeness.}
    \label{fig: time-DEPENDENT gamma2 opt}
\end{figure}

Several configurations have been tested (see App.~\ref{appendix: numerical opt, time-dependent}). Eventually, the most promising setup involves a square-function $\gamma_1(t)$ and two-step functions $\gamma_2(t)$ and $\gamma_3(t)$, described as
\begin{equation}
    \label{eq: time-dep gammas, MAIN}
    \begin{split}
    \gamma_1(t) &= \gamma_1\Theta(t-t_0)\Theta(t_a^{(s)} - t) \\
    \gamma_2(t) &= \gamma_2\Theta(t-t_a^{(s)}) \\
    \gamma_3(t) &= \gamma_3\Theta(t_a^{(c)} - t)  \, ,
    \end{split}
\end{equation}
where $\Theta(t-t_0)$ represents the Heaviside function. In the simulations, the arrival time of the control photon has been set to $\omega_0 t_a^{(c)} =4.3$ (i.e., $t_a^{(c)} =1.075$~$\mu$s), while $\omega_0 t_0 = 3.7$ (that is, $t_0 = 0.925$~$\mu$s) corresponds to the the square function $\gamma_1(t)$ turning on. We notice that $t_a^{(c)}$ has been slightly anticipated with respect to the previous optimizations; on the other hand, the coefficients $\gamma_i$ in Eqs.~(\ref{eq: time-dep gammas, MAIN}) are taken as the optimized time-independent coupling parameters found above. In such a configuration, the photon flux difference integrated in steady state (i.e., in the limit of long integration time, $t\to \infty$) gives $\Phi_2^{(s,c)} - \Phi_2^{(c)} \simeq 0.489$. Although a significant contribution to the flux is due to the control photon, a conspicuous fraction can be attributed to the signal photon itself (as it can be inferred from Fig.~\ref{fig: time-DEPENDENT gamma2 opt}).

We can now analyze the realistic switching capability of the device by addressing model B, which involves the additional filtering cavity. In this case, the output channel of the network coincides with the output of Cavity 2, which is weakly coupled to Cavity 1 through the interaction strength $J$. 

In particular, by using the same strategy outlined above, we numerically found that $J=1$ MHz is an optimal value, leading to the steady state figure of merit $\Phi_2^{(s,c)} - \Phi_2^{(c)} \simeq 0.429$. With the aim of highlighting the switching effect of the device, we hereby compare the output flux calculated for the two alternative cases characterizing the switch behavior in model B: the switch-ON and the switch-OFF modes. First, both signal and control photons are present, and the calculated flux is $\Phi_2^{(s,c)}$, which corresponds to the switch-ON configuration. Then, we assume that only the signal photon drives the system, while the control pulse is absent: the corresponding output flux is $\Phi_2^{(s)}$, and the device is in switch-OFF mode. In addition, we also show the situation when only the control photon is present, by displaying $\Phi_2^{(c)}$. 
The difference between models A and B clearly emerges by comparing Figs.~\ref{fig: switch process}(a) and \ref{fig: switch process}(b): while in Fig.~\ref{fig: switch process}(a) the steady-state output flux due to the control photon is still relevant ($\Phi_2^{(c)} \simeq 0.549$), in Fig.~\ref{fig: switch process}(b) it is strongly suppressed ($\Phi_2^{(c)} \simeq 10^{-4}$). The control photon contribution  is therefore negligible in model B, meaning that the flux contribution $\Phi_2^{(s,c)} \simeq 0.429$ due to the signal photon when the switch is in ON mode is, practically speaking, identical to the difference $\Phi_2^{(s,c)} - \Phi_2^{(c)}$. 
{It is evident that, while in Fig.~\ref{fig: switch process}(a) both the signal and the control photons contribute to the overall output flux, in Fig.~\ref{fig: switch process}(b) the output flux is almost entirely determined by the signal photon pulse, due to the presence of the filtering cavity that is off-resonant with respect to the control photon. We also notice that in both cases no output from the signal photon is detected, if the control photon is not present. This  behavior shows the effective switching capability of the device when operated in the model B configuration, as initially illustrated in Sec.~\ref{sec:switch_description}: the signal photon is efficiently transmitted to the output port if the control photon is present, while no appreciable transmission is detected if the control is absent. In particular, we notice that in the latter condition the integrated flux goes to $\Phi_2^{(s)} \simeq 1.22 \cdot 10^{-6}$ in steady state, which is five orders of magnitude smaller than the total flux contribution in presence of both signal and  control photons, denoting an excellent switching contrast for classical operations. 

\begin{figure}[t]
\centering
\includegraphics[width=0.48\textwidth]{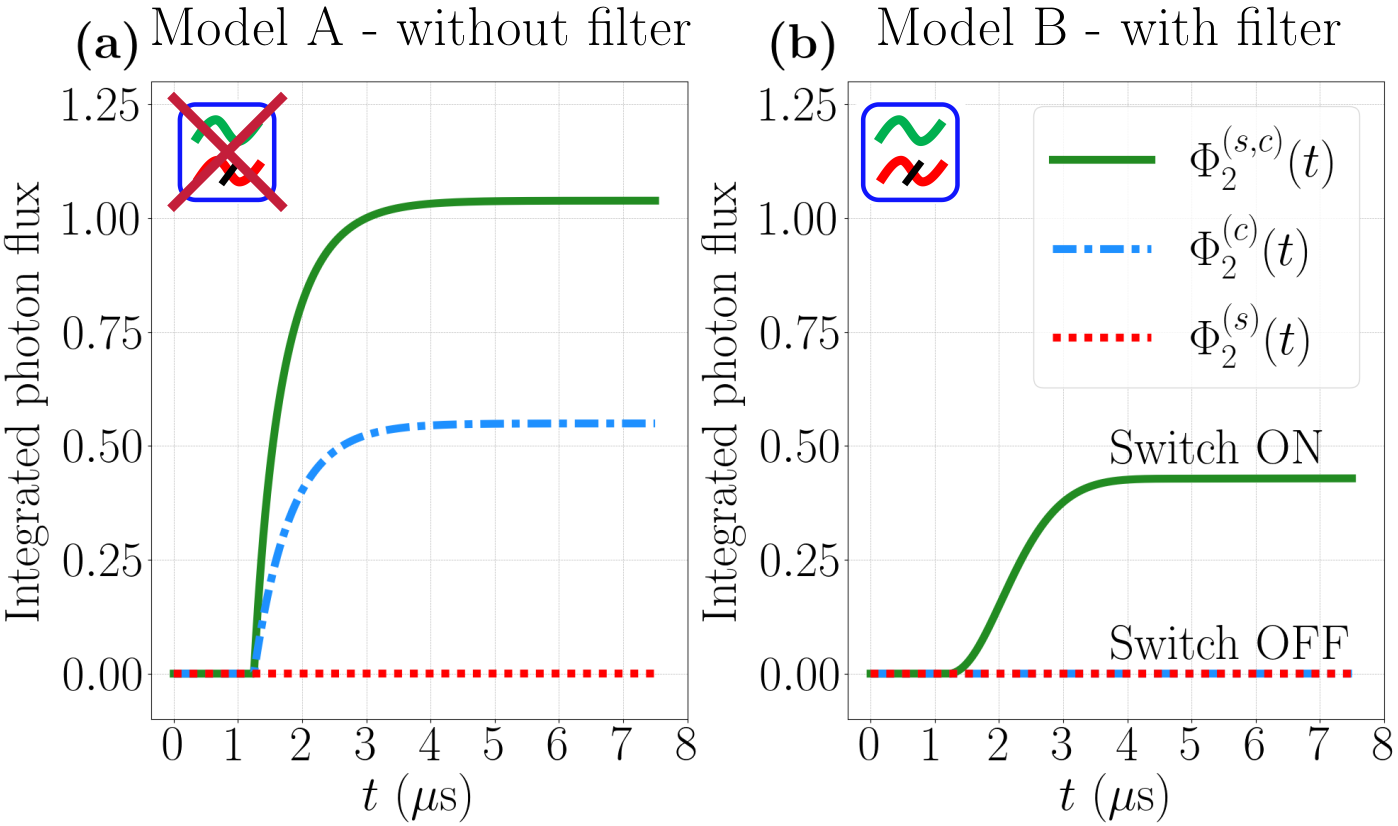}
\caption{\justifying Comparison between the photon fluxes $\Phi_2^{(s,c)}(t)$, $\Phi_2^{(s,c)}(t)$, and $\Phi_2^{(s,c)}(t)$ integrated  up to time $t$, as a function of $t$, obtained for the optimized configurations of (a) model A and (b) Model B, respectively. The insets in both panels represent the switch symbol: model A is not working as a switch for the signal photon, while model B is an operating single-photon switch. The results shown in panel (a) are the same as the ones already reported in Fig.~\ref{fig: time-DEPENDENT gamma2 opt}, and are reported here for completeness and direct comparison to panel (b). }
\label{fig: switch process}
\end{figure}

In the literature, various figures of merit have been presented to actually quantify the efficiency of a switching device, and especially the difference between the ON- and the OFF-condition. Specifically referring to the literature related to single-photon switching in similar setups,  a \textit{switching contrast} $\mathcal{C}$ has been defined, e.g., in Ref.~\cite{StolyarovPRA2020}, quantified as the difference between two steady state transmission probabilities: the probability of transmitting a single photon when the control qubit is in the ground state, and the one for the case in which the qubit is in its excited state. In the same spirit, a switching contrast $C_s$ is also defined, e.g., in Ref.~\cite{hartmann2013single}. 
In addition, a measure of the discrepancy between the ON/OFF conditions has been proposed: in Ref.~\cite{hartmann2013single} a ratio between probabilities is employed, as well as in Ref.~\cite{Wang2022}. In the latter, the extinction ratio is also defined as $R = 10\log_{10}(n_{\ket{1}}^{\text{open}} / n_{\ket{0}}^{\text{open}})$,} typically given in dB. In the latter expression, $n_{\ket{N}}^{\text{open}}$ is the average number of output photons in the presence of a single gate photon (i.e., $\ket{N}=\ket{1}$) or in its absence  (i.e., $\ket{N}=\ket{0}$). Here, we analogously define the extinction ratio as 
\begin{equation}
    \label{eq: extinction ratio}
    R(t) = 10\log_{10}\biggl( \frac{\Phi_2^{(s,c)}(t)}{\Phi_2^{(s)}(t)} \biggr) \, ,
\end{equation}
in which  $\Phi_2^{(s,c)}(t)$ and $\Phi_2^{(s)}(t)$ have been defined before. The calculated ratio in Eq.~\ref{eq: extinction ratio} for the switching process depicted in Fig.~\ref{fig: switch process}(b) is shown in Fig.~\ref{fig: extinction ratio}. {In order to compare our results with the literature, we calculate the steady-state extinction ratio $R_{\infty} \equiv \displaystyle\lim_{t\rightarrow +\infty} R(t)$. Notably, in our setup the steady-state extinction ratio reaches a value of $R_\infty \simeq 56.14$ dB}, which represents a significant improvement, albeit theoretical, over the values originally obtained in the proposal of Ref.~\cite{hartmann2013single}, as well as in recent experiments~\cite{Wang2022}. In addition, we have used realistic experimental parameters to achieve this result, which shows the effectiveness of our proposal and its potential interest for actual implementations in superconducting circuits.

\begin{figure}
\centering
\includegraphics[width=0.45\textwidth]{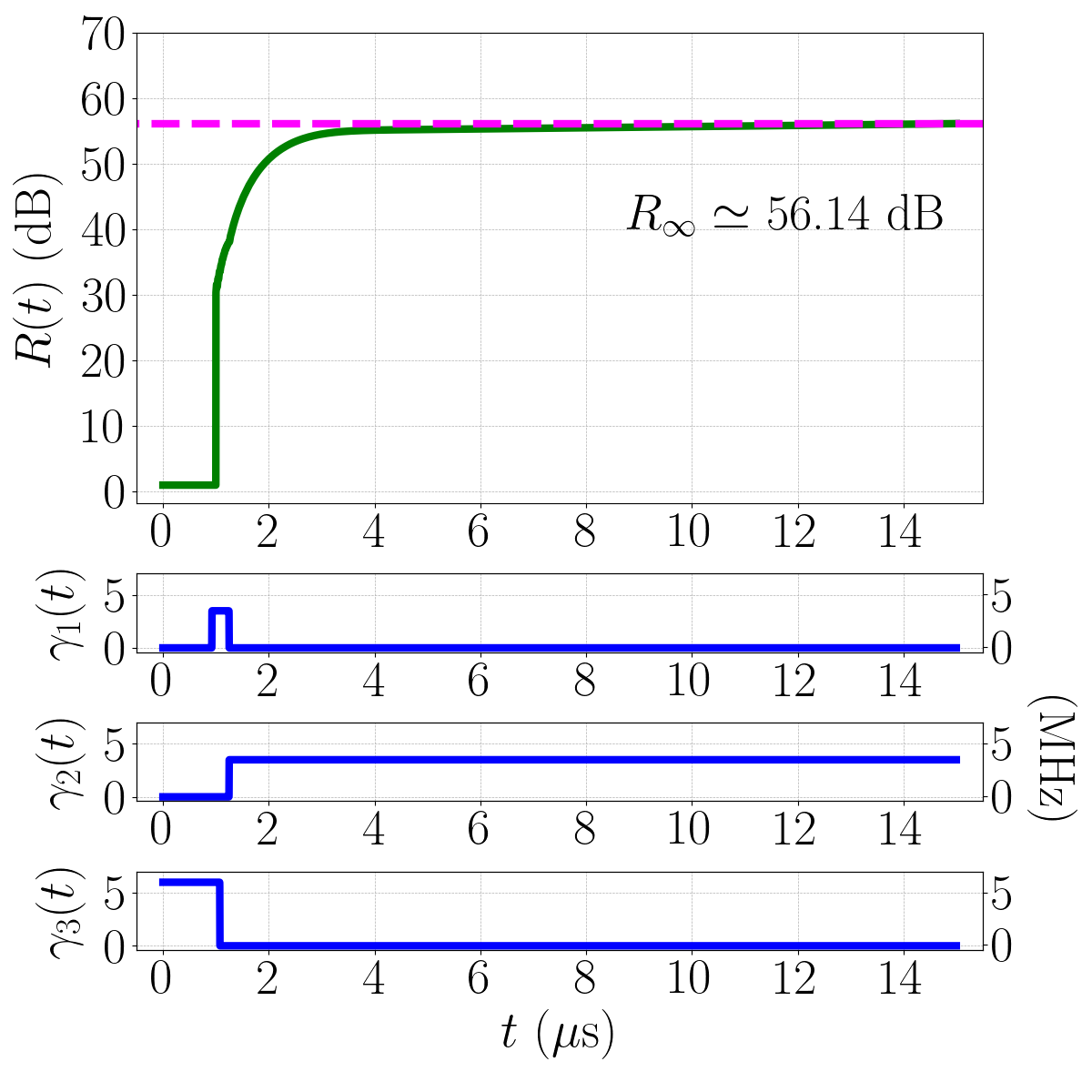}
\caption{\justifying Extinction ratio, as defined in Eq.~(\ref{eq: extinction ratio}), as a function of time and calculated for the switching process described within the scheme defined as model B, see Fig.~\ref{fig: switch process}(b). The steady state extinction ratio is $R_{\infty}\simeq 56$ dB. For completeness, the time-dependent coupling rates $\gamma_i(t)$ from Eqs.~(\ref{eq: time-dep gammas, MAIN}) are explicitly plotted in the lower panels.}
\label{fig: extinction ratio}
\end{figure}

\section{Discussion}

\begin{figure}
\centering
\includegraphics[width=0.5\textwidth]{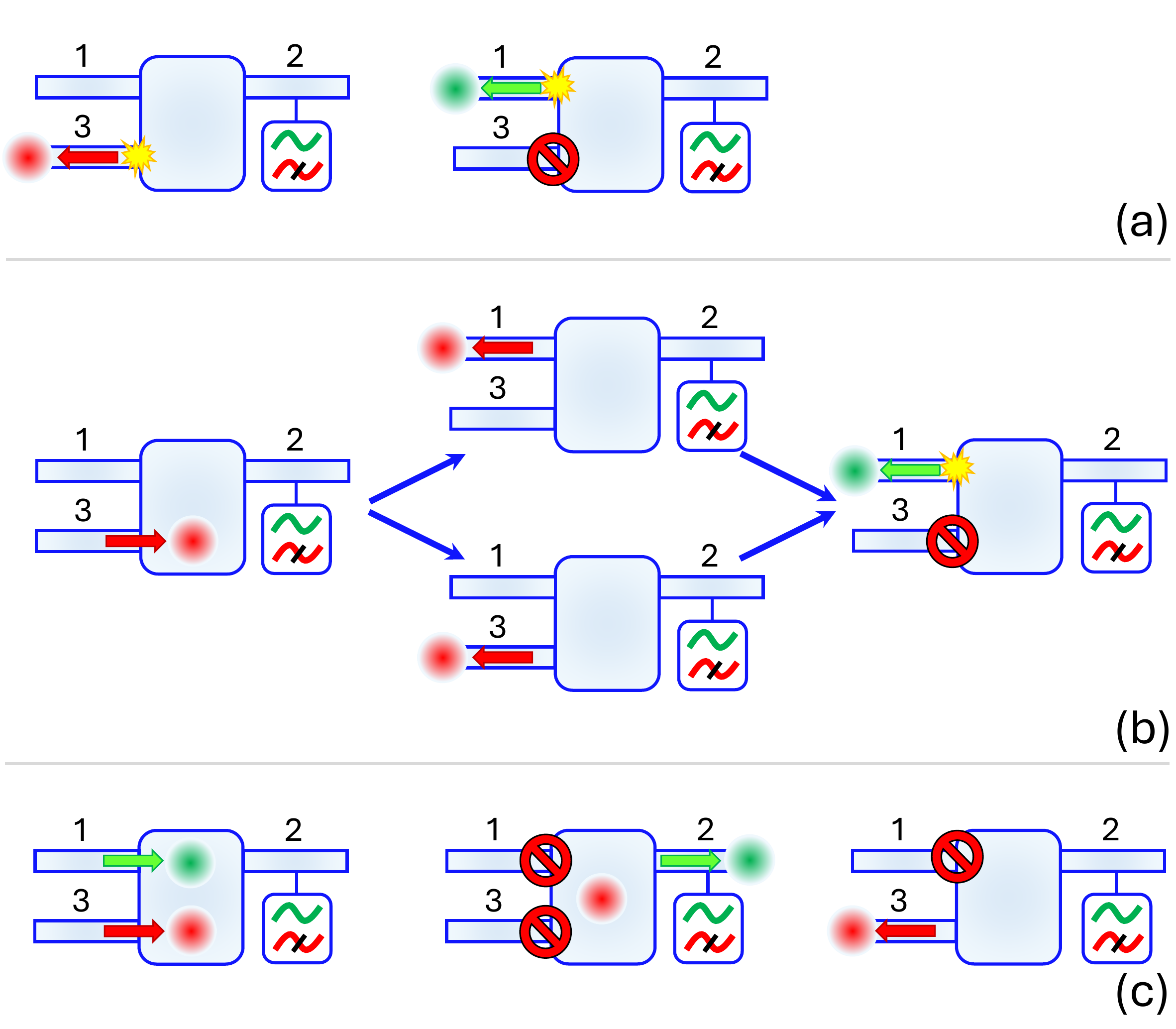}
\caption{\justifying Summary of the possible situations that may occur in a single-photon switching process. (a) The control photon is back-scattered from the system, and the signal photon is also reflected as a consequence. (b) The control photon enters the system, but it escapes from output channels 1 or 3 before the  signal photon arrives; the latter is then back-scattered. (c) Both control and signal photons enter the system; thus, the signal can escape from output channel 2, while the control photon is trapped inside the JC sub-system. In the latter case, the output channel 3 could be re-opened after a sufficiently long time ($t_1 \sim 30\text{ }\mu$s), allowing the control photon to be re-emitted, and eventually recovered.}
\label{fig: fig7}
\end{figure}

Here we discuss an additional feature that might be relevant for technological applications of the single-photon switching device proposed in this work: what happens to the control photon after the switching operation on the signal has been performed?
In fact, the possibility of time-modulating the coupling factors, $\gamma_i(t)$, naturally provides the opportunity to retrieve the control photon after the process. Indeed, this photon can be recovered by waiting enough time to allow for the switching process to be completed (e.g., in the order of $t_1 \sim 7\text{ $\mu$s}$). As shown in Fig.~\ref{fig: fig7}, within this time frame the control photon could be either (a) back-scattered from the system, or (b) absorbed by the system and re-emitted before the turning-off of $\gamma_1(t)$ and $\gamma_3(t)$, or even (c) absorbed by the system and trapped for a longer time. 
In case (a), the control photon may be recovered after escaping from channel 3. Case (b) is the most delicate: if the control photon escapes from channel 3, it could be retrieved, but if it goes out of channel 1 it would be lost. However, such an event could be partly inhibited by exploiting a tunable coupling factor from channel 1, $\gamma_1(t)$. 
In the most desirable situation, i.e., case (c), the control photon remains inside the joint system (Cavity 1 + qubit), ready to be used for a subsequent switching operation on the next single photon: $\gamma_3(t)$ could be turned on after time $t_1$, such that the control photon be emitted by the system through the only available channel, i.e., channel 3 itself. Furthermore, while in case (c) the control photon would be recovered only after time $t_1$, in case (a) - and, only in part of the processes, in case (b) - the control photon should be detected well before time $t_1$. This could be a way to determine whether the switching process has actually taken place or not, even without a direct measurement of the signal photon. 

In the perspective of implementing a network of cascaded devices, the presence of a possible backaction after the photon measurement must be taken into consideration. Indeed, such an issue could be significantly relevant
depending on how strongly the measuring instrument is coupled with the measured systems. Nonetheless, if the instrument is only perturbatively coupled to the device via a sufficiently low interaction strength, it can be expected that the switch performances will not be significantly affected by  backaction of the output photon measurements. In general, it is worth stressing that a proper account of the backaction in a full quantum network could only be achieved by including both measuring and measured systems into the SLH triples, which is beyond the scopes of the present work. \\
A further issue, that might be relevant when a sequence of single photon signals with high repetition rate has to be successfully switched, is the overall process duration. In fact, according to our preliminary analysis (not shown), the overall switching time is found to mostly depend on the coupling strength, $J$. In particular, we found that it should be possible to reduce the process duration; however, this comes at the cost of a reduced efficiency. A quantitative analysis of these aspects goes beyond the present work, and it might be presented elsewhere.

We finally place our work in the context of the existing literature. As mentioned, the original idea of using a scheme based on the first two manifold of the anharmonic JC spectrum to favor single-photon switching has already been put forward and experimentally investigated in the optical domain, i.e., employing a single semiconductor quantum dot coupled to a photonic crystal cavity in a pump/probe configuration~\cite{volz2012ultrafast} .  However, here we exploit the rich degree of tunable parameters that can be realized in the microwave domain through superconducting circuit setups to generalize the scheme proposed in \cite{volz2012ultrafast}. In particular, we propose to exploit the possibility of separately driving the qubit and the cavity with control and signal photons, respectively. The possibility of implementing such a model in a superconducting cavity QED setup has also been mentioned in Ref.~\cite{manzoni2014}, as an alternative to a three-level system model; however, this idea was not further investigated. We also notice that a similar setup to the one discussed here 
has already been analyzed in Ref.~\cite{xu2016single}, but working in a dispersive regime~\cite{schuster2007resolving} (i.e., for $|\Delta| \gg g$), while our model refers to the resonant strong coupling regime ($\Delta = 0$) to maximize the single-photon nonlinearity. On the other hand, the model proposed in Ref.~\cite{hartmann2013single} relies on a similar idea of separately driving two coupled oscillators, two qubits in that case, thus not producing a JC spectral response. In fact, in our case the presence of a cavity and a qubit originates a fully asymmetrical model, which allows to better distinguish and discriminate between the signal and the control pulses. It should be noted that none of the previous theoretical proposals seem to have been realized in state-of-art superconducting circuit technology, to the best of our knowledge. We then envision significant room for development in this field, with the ultimate target of efficiently modulating single microwave photons with single and eventually recoverable control photons.

\section{Conclusions}

We have applied the SLH formalism to analyze an elementary quantum optical model that can be used to implement a superconducting switch for single microwave photons, to be employed, e.g., in classical signal processing exploiting advances in quantum technologies. 

The key building block (model A) consists of a microwave resonator coupled to a single superconducting qubit in the strong coupling regime. A further coupled cavity can be added, to be used as a filter to purify the output signal (model B). Referring to model A, we exploit the Jaynes-Cummings spectral anharmonicity at the single excitation level to prevent or allow the transmission of a single (signal) photon through the device, determined by the absence or by the presence of a (single) control photon. We notice that in this case  the absence of any filter implies that the control photon escapes from the output channel together with the signal photon. We therefore extend our theoretical analysis to model B, by adding the filtering cavity. We have shown that in this situation the elementary quantum network becomes a realistic single-photon switch, by fully suppressing transmission of the control photon, while only the signal photon can be released from the device in the ON operational mode. 

We have performed extensive numerical simulations to optimize the model parameters  to achieve the best efficiency in terms of switching capability. In particular, we initially maximized the probability of the signal and control photons to be simultaneously absorbed in model A, as well as the integrated photon flux at the output port of the device. The latter quantity provides a direct and effective figure of merit to compare our theoretical predictions with possible experimental outcomes. In this context, we assumed time-independent coupling rates between the network and the input-output waveguides. \\
As a next step, we generalized to piecewise constant time-dependent parameters, in which the constant values are assumed as the coupling rates optimized before. We thus analyzed different configurations in order to further increase the amount of output photon flux. Finally, we numerically solved for the full model B, showing that the device indeed behaves as an effective single-photon switch. 
In particular, our SLH simulations have evidenced that an optimal steady state output flux, $\Phi_2^{(s,c)}\simeq 0.429$ (see Fig.~\ref{fig: switch process}) can be achieved within a $7\text{ $\mu$s}$ timescale. We also noticed how the switch OFF mode of this scheme is far more efficient, allowing for an almost complete extinction in the absence of a single control photon. Such a switching contrast is quantitatively described by the extinction ratio defined in Eq.~(\ref{eq: extinction ratio}), which can reach $R_{\infty}\simeq 56$ dB for our optimal setup configuration (Fig.~\ref{fig: extinction ratio}). 

Finally, we have discussed the possibility that such a scheme allows the recovery of the control photon, which is achieved by properly tuning the coupling factors, $\gamma_i(t)$, during the switching process. This is promising in view of an actual application of this setup for classical signal processing in the microwave domain, which would allow to minimize the loss of information. We envision that a proof-of concept demonstration of this device could be realized in state-of-art superconducting quantum circuits, currently developed for quantum information processing, and possibly open new research directions.

\section*{Acknowledgements}

This research was supported by the Italian Ministry of Research (MUR) through the PNRR project PE0000023 - National Quantum Science Technology Institute (NQSTI). The authors warmly acknowledge D. Bajoni for scientific discussions.


%

\newpage

\appendix

\section{Theoretical background}\label{appendix: theory in details}

The theoretical analysis performed in this work exploits the input-output theory, ultimately underlying the SLH formalism. The key steps consist of (a) the definition of the total Hamiltonian, (b) the definition of the itinerant fields, (c) the assumption of a few key hypotheses, and (d) switching to the interaction picture. 
On these premises, we can therefore build a proper SLH scheme of the device, and derive the related differential equations for the systems density matrix, to be solved numerically. Once solved for the density matrix, it is then straightforward to compute the expectation values for all the quantities of interest as a function of time.

The setup for single-photon switching presented in this work analyzes the case of two quantum harmonic oscillators (Cavity 1 and Cavity 2) and a single two-level system (the qubit). Cavity 1 and the qubit are assumed to be direcly interacting between each other, while Cavity 2 is an additional feature considered in a second step, and it is assumed to be coupled only to cavity 1. The usefulness of this second harmonic oscillator emerges when it is necessary to filter the output of the device, and thus we refer to it as the \textit{filtering cavity}. We will now detail all the elements of the theoretical model employed to describe this system. In the next Section, we will summarize the SLH formalism, which is the method employed to solve the coupled differential equations deriving from this model.

\subsection{System-field coupling}
We start by assuming (i) a linear, (ii) weak and (iii) uniform (over a broad range of field frequencies) coupling between the localized systems and the environment. Hypothesis (ii) is known as the \textit{Born approximation}, while hypothesis (iii) is a frequency-domain version of the \textit{Markov approximation}. In order to make use of the SLH formalism, we also assume that the medium in which the photons travel is dispersionless. Furthermore, we assume that we the \textit{rotating wave approximation} (RWA) for the system-field coupling is satisfied. 

In the present work, we have distinguished between a ``Model A'' and a ``Model B'', whose Hamiltonian descriptions are given in Eqs.~(\ref{eq: Hamiltonian-A})-(\ref{eq: bath-sys interaction Hamiltonians}). Here, we provide the essential details about the system-environment coupling, from which the definition of coupling rates, $\gamma_i(t)$, derives. The Hamiltonian terms describing the interaction between the systems and the travelling fields are given in Eq.~(\ref{eq: H bath-sys A}) for ``Model A'', and Eq.~(\ref{eq: H bath-sys B}) for ``Model B'', with the following formal expressions 
\begin{equation}
\label{eq: Interaction Hamiltonians system-field, appendix}
\begin{split}
    &\hat{H}^{(\text{cavity 1, } s)}_{\text{int}} = -i\hbar\int_0^{+\infty} \kappa_1 \bigl[ \bigl(\hat{a}_1 + \hat{a}_1^\dagger \bigr) \bigl(\hat{b}_s - \hat{b}^\dagger_s \bigr)\bigr]d\omega \\
    &\hat{H}^{(\text{cavity 1, } o)}_{\text{int}} = -i\hbar\int_0^{+\infty} \kappa_2 \bigl[ \bigl(\hat{a}_1 + \hat{a}_1^\dagger \bigr) \bigl(\hat{b}_o - \hat{b}^\dagger_o \bigr)\bigr]d\omega \\
    &\hat{H}^{(\text{cavity 2, } o)}_{\text{int}} = -i\hbar\int_0^{+\infty} \kappa_2 \bigl[ \bigl(\hat{a}_2 + \hat{a}_2^\dagger \bigr) \bigl(\hat{b}_o - \hat{b}^\dagger_o \bigr)\bigr]d\omega \\
    &\hat{H}^{(\text{qubit, } c)}_{\text{int}} = -i\hbar\int_0^{+\infty} \kappa_3 \bigl[ \bigl(\hat{\sigma}_- + \hat{\sigma}_+ \bigr) \bigl(\hat{b}_c - \hat{b}^\dagger_c \bigr)\bigr] d\omega \, ,
\end{split}
\end{equation}
in which $\hat{b}_p = \hat{b}_p(\omega)$, with $p=s,c,o$ such that $s =$ signal, $c = $ control, $o = $ output. In this description, $\kappa_1 = \kappa_1(\omega)$ is the coupling rate between Cavity 1 and the field mode associated with the input-output channel 1; $\kappa_3 = \kappa_3(\omega)$ is  the qubit-field coupling strength, and it is associated with channel 3. Within ``Model A'', the interaction rate $\kappa_2 = \kappa_2(\omega)$ describes the coupling between Cavity 1 and the output field mode; yet, $\kappa_2(\omega)$ in ``Model B'' represents the coupling between the same output field mode and Cavity 2. 

After changing reference frame (see subsec.~\ref{subsec: change of reference frame}), 
we define $\hat{b}_s(t) = 1/2\pi\int d\omega \hat{b}_s(\omega)e^{i(\omega-\omega_s)t}$, $\hat{b}_o(t) = 1/2\pi\int d\omega \hat{b}_o(\omega)e^{i(\omega-\omega_o)t}$ and $\hat{b}_c(t) = 1/2\pi\int d\omega \hat{b}_c(\omega)e^{i(\omega-\omega_c)t}$, as in Refs.~\cite{Baragiola:2014, GardinerZoller, Combes_SLH_framework}. 
Actually, the latter are input itinerant fields, i.e., photon wavepackets that travel towards the localized systems. Their frequencies define the modes of the electromagnetic field that are coupled to the localized systems: these can be excited (and, viceversa, relax) by absorbing (or emitting) one or more photons in these field modes. In that sense, these field modes represent the environment to which the localized systems are coupled. We also notice that, in this work, the output carrier frequency,  $\omega_o$, coincides with the input signal carrier frequency, i.e., $\omega_o \equiv \omega_s$. \\
Under the hypotheses above, the Hamiltonian terms describing interactions between the localized systems and the corresponding propagating fields become
\begin{equation}
    \label{eq: Interaction Hamiltonians, system - fields, appendix}
    \begin{split}
    &\hat{H}^{(\text{cavity 1, } s)}_{\text{int}} = i\hbar\sqrt{\gamma_1}[\hat{a}_1\hat{b}_s^\dagger(t) - \hat{a}_1^\dagger \hat{b}_s(t)] \\
    &\hat{H}^{(\text{cavity 1, } o)}_{\text{int}} = i\hbar\sqrt{\gamma_2}[\hat{a}_1\hat{b}_o^\dagger(t) - \hat{a}_1^\dagger \hat{b}_o(t)] \\
    &\hat{H}^{(\text{cavity 2, } o)}_{\text{int}} = i\hbar\sqrt{\gamma_2}[\hat{a}_2\hat{b}_o^\dagger(t) - \hat{a}_2^\dagger \hat{b}_o(t)] \\
    &\hat{H}^{(\text{qubit, } c)}_{\text{int}} = i\hbar\sqrt{\gamma_3}[\hat{\sigma}_-\hat{b}_c^\dagger(t) - \hat{\sigma}_+ \hat{b}_c(t)] \, .
    \end{split}
\end{equation}
 Here, the time-independent coefficients $\gamma_i$ describe  the coupling rates between one of the three systems and the $i$-th channel. They originate from the coupling strengths $\kappa_i$, as a consequence of the Markov approximation. In this very case, indeed, $\gamma_i = 2\pi|\kappa(\omega_i)|^2$, with $\omega_1 \equiv \omega_s$, $\omega_2 \equiv \omega_o$, and $\omega_3 \equiv \omega_c$.

\subsection{Single-photon wavepackets}
We now consider the problem of selecting the proper temporal wave forms for the single-photon wave packets. First, we notice that when a two-level system is excited, its spontaneous decay is described by a lowering exponential probability function of time, i.e., $p(t)\sim e^{-k(t-t_0)}$. Therefore, the corresponding single-photon wavepacket generated from this decay process has a rising exponential temporal dependence, $f(t)\sim e^{+k(t-t_0)}$. However, it is not possible to achieve an optimal absorption by a two-level system, if the incoming photon has a such a wavepacket shape. On the contrary, we can obtain a nearly maximal (i.e., $\sim 1$) absorption probability if the photon radiation state corresponds to the \textit{time reversed} of such a spontaneously-emitted photon \cite{stobinska2009perfect, wenner2014catching}. This means that the incoming wavepacket should have a lowering exponential shape, i.e., $f(t)\sim e^{-k(t-t_0)}$, as shown in Fig.~\ref{fig: fig1B}. 

Therefore, based on these earlier studies we directly choose lowering exponential single-photon wavepackets in our simulations, in order to maximize the absorption probability. We then define the \textit{time reversed} wavepacket $\xi_p(t)$ as follows:  
\begin{equation}
    \label{eq: time-reversed wavepacket, appendix}
    \xi_p(t) = \begin{cases}
			\sqrt{\Delta\omega_p} e^{\frac{\Delta\omega_p}{2}(t-t_a^{(p)})} & \text{if $t < t_a^{(p)}$}\\
            0 & \text{if $t \geq t_a^{(p)}$} \, ,
		 \end{cases}
\end{equation}
in which $p = s,c$ denote the signal and the control pulses, respectively. Actually, $\xi_p(t)$ is the function that is employed in the differential equations originated from the SLH formalism, although it corresponds to the time reversal of the incoming photon wavepackets, which are lowering exponential functions of time, in fact.  

\subsection{Change of reference frame}
\label{subsec: change of reference frame}
Before taking advantage of the SLH framework, it is necessary to switch to the interaction picture \cite{Combes_SLH_framework, Baragiola:2014}. In order to do so, it is useful to define the following two rotation operators
\begin{equation}
    \label{eq: rotations A and B}
    \begin{split}
    \hat{\mathcal{R}}_A(t) = \text{exp}\{& -i\omega_s\hat{a}_1^\dagger \hat{a}_1 t -i\omega_c\hat{\sigma}_+ \hat{\sigma}_-t \\
    &- \frac{i}{\hbar} \hat{H}_{\text{free}}^{(s)}t - \frac{i}{\hbar} \hat{H}_{\text{free}}^{(o)}t - \frac{i}{\hbar}\hat{H}_{\text{free}}^{(c)}t\} \\
    \hat{\mathcal{R}}_B(t) = \text{exp}\{& -i\omega_s\hat{a}_1^\dagger \hat{a}_1 t -i\omega_s\hat{a}_2^\dagger \hat{a}_2 t  -i\omega_c\hat{\sigma}_+ \hat{\sigma}_-t \\
    &- \frac{i}{\hbar} \hat{H}_{\text{free}}^{(s)}t - \frac{i}{\hbar} \hat{H}_{\text{free}}^{(o)}t - \frac{i}{\hbar}\hat{H}_{\text{free}}^{(c)}t\} \, ,
    \end{split}
\end{equation}
which will be applied to ``Model A'' and ``Model B'', respectively. These allow us to perform a change of reference frame, since we can rewrite the every Hamiltonian with the free evolution energies referred to the single-photon carrier frequencies, i.e., $\omega_s$ and $\omega_c$ in our case. 

Let us consider the Hamiltonian terms defined in Eqs.~(\ref{eq: Hamiltonian-A})-(\ref{eq: bath-sys interaction Hamiltonians}): ``Model A'' is described by $\hat{H}_{\text{tot}}^A = \hat{H}_A + \hat{H}^{\text{sys-env}}_A$, while as ``Model B'' is represented by $\hat{H}_{\text{tot}}^B = \hat{H}_B + \hat{H}^{\text{sys-env}}_B$. If we apply Eqs.~(\ref{eq: rotations A and B}), the transformed operators read
\begin{equation}
    \label{eq: Rotated Hamiltonian A, B - general}
        \tilde{\hat{H}}_{\text{tot}}^X = \hat{\mathcal{R}}_X^\dagger(t) \hat{H}_{\text{tot}}^X \hat{\mathcal{R}}_X(t)  \, ,
\end{equation}
with $X=A,B$. Equivalently, the density matrix of the localized systems can be expressed as $\tilde{\hat{\rho}}_X(t) = \hat{\mathcal{R}}_X(t) \hat{\rho}_X \hat{\mathcal{R}}_X^\dagger(t)$, which satisfies the following evolution equation:
\begin{equation}
    \label{eq: Liouville equation for rho rotated}
    \frac{d}{dt}\tilde{\hat{\rho}}_X(t) \equiv -\frac{i}{\hbar}\biggl[ \hat{H}_{\text{eff}}^X, \tilde{\hat{\rho}}_X(t)\biggr] \, ,
\end{equation}
for a specific effective system Hamiltonian $\hat{H}_{\text{eff}}^X$. Within this notation, the Hamiltonian is derived from Eq.~(\ref{eq: rotations A and B}) as
\begin{equation}
    \label{eq: Heff}
\hat{H}_{\text{eff}}^X \equiv \tilde{\hat{H}}_{\text{tot}}^X(t) + i\hbar\hat{\mathcal{R}}_X^\dagger(t)\frac{d}{dt}\hat{\mathcal{R}}_X(t) \, .
\end{equation}
We finally get the following explicit expressions for the effective Hamiltonians of ``Model A'' and ``Model B''
\begin{equation}
    \label{eq: Rotated EFFECTIVE Hamiltonians}
    \begin{split}
        \hat{H}_{\text{eff}}^A &=  \hat{H}_{\text{sys}}^A + \tilde{\hat{H}}^{\text{sys-env}}_A \\
        &= -\hbar\Delta_s \hat{a}_1^\dagger\hat{a}_1 -\hbar\Delta_c \hat{\sigma}_+\hat{\sigma}_- + \tilde{\hat{H}}_{\text{int}}^{(\text{cavity 1, qubit})} \\ &\quad+ \tilde{\hat{H}}^{\text{sys-env}}_A \\
        \hat{H}_{\text{eff}}^B &=  \hat{H}_{\text{sys}}^B + \tilde{\hat{H}}^{\text{sys-env}}_B \\
        &= -\hbar\Delta_s \hat{a}_1^\dagger\hat{a}_1 -\hbar\Delta_c \hat{\sigma}_+\hat{\sigma}_- -\hbar\Delta_s \hat{a}_2^\dagger\hat{a}_2 \\
        &\quad+ \tilde{\hat{H}}_{\text{int}}^{(\text{cavity 1, qubit})} + \hat{H}_{\text{int}}^{(\text{cavity 1, cavity 2})} \\
        &\quad+ \tilde{\hat{H}}^{\text{sys-env}}_B \, ,
    \end{split}
\end{equation}
in which $\Delta_s = \omega_s - \omega_0$ and $\Delta_c = \omega_c - \omega_0$. Besides, the  interaction Hamiltonian between Cavity 1 and qubit can be expressed in this rotated frame as
\begin{equation}
    \label{eq: rotated H_int Cavity 1 - Qubit}
        \tilde{\hat{H}}_{\text{int}}^{(\text{cavity 1, qubit})} = \hbar g [\hat{a}_1\hat{\sigma}_+ e^{-i\Delta_{s,c}t} + \hat{a}_1^\dagger\hat{\sigma}_- e^{+i\Delta_{s,c}t}]  \, ,
\end{equation}
in which $\Delta_{s,c} = \omega_s - \omega_c$ denotes the detuning between the signal and control carrier frequencies, respectively. We also notice that the interaction Hamiltonian between Cavity 1 and Cavity 2 is invariant under such transformation: $\tilde{\hat{H}}_{\text{int}}^{(\text{cavity 1, cavity 2})} = \hat{H}_{\text{int}}^{(\text{cavity 1, cavity 2})}$. Also, the system-field interacting terms are expressed exactly as in Eqs.~(\ref{eq: Interaction Hamiltonians, system - fields, appendix}), obtained by defining the input-output operators $\hat{b}_p(t)$ after the change of reference frame. \\
As a final remark, the Hamiltonian terms describing the evolution of the localized systems are obtained as
\begin{equation}
    \label{eq: SYSTEM Hamiltonians}
    \begin{split}
        \hat{H}_{\text{sys}}^A &= -\hbar\Delta_s \hat{a}_1^\dagger\hat{a}_1 -\hbar\Delta_c \hat{\sigma}_+\hat{\sigma}_- + \tilde{\hat{H}}_{\text{int}}^{(\text{cavity 1, qubit})} \\
        \hat{H}_{\text{sys}}^B
        &= -\hbar\Delta_s \hat{a}_1^\dagger\hat{a}_1 -\hbar\Delta_c \hat{\sigma}_+\hat{\sigma}_- -\hbar\Delta_s \hat{a}_2^\dagger\hat{a}_2 \\
        &\quad+ \tilde{\hat{H}}_{\text{int}}^{(\text{cavity 1, qubit})} + \hat{H}_{\text{int}}^{(\text{cavity 1, cavity 2})} \, .
    \end{split}
\end{equation}
We stress that the latter play a prominent role in the SLH framework, as discussed in the following subsection.

\section{The SLH framework of input-output quantum networks}\label{appendix: slh framework}

We hereby summarize the SLH framework and its extensions, for completeness. For an extensive review and discussion, we refer to the excellent Ref.~\cite{Combes_SLH_framework}. Essentially, the SLH is a theoretical tool aimed at simulating open quantum systems in an input/output configuration. Ultimately, its theoretical roots trace back to the \textit{quantum stochastic calculus} and the \textit{input-output theory} \cite{Charmichael, GardinerZoller, GardinerCollett}, generalized to become an invaluable method to treat complex quantum networks in a modular way.

\subsection{SLH formalism}

The quantum description of a localized system  (e.g., a two-level system, or an optical cavity) interacting with one (or more) itinerant field mode (i.e., time-dependent photon wavepackets) can be formulated as an input-output theory with a system-environment interaction. As an initial state, it can be assumed that the environment is the vacuum state $\ket{0}$. Most importantly, we assume that the hypotheses mentioned in App.~\ref{appendix: theory in details} are valid, together with the additional assumption that there is \textit{no back-scattering} between cascaded quantum systems (although this is not necessary for the purposes of this paper). \\
Within the SLH formalism, an operators \textit{triple} $\mathbf{G} = (\mathbf{S}, \mathbf{L}, \hat{H})$ is assigned to the localized system. Here, $\mathbf{S}$ generally represents a matrix of scattering operators, $\mathbf{L}$ is a vector of coupling operators, and $\hat{H}$ is the system Hamiltonian in interaction picture, i.e., rotated via a unitary transformation such as Eqs.~(\ref{eq: rotations A and B}). While the triple describes the evolution of the system, $\mathbf{S}$ takes into account its scattering with between the itinerant field modes, and $\mathbf{L}$ contains the information regarding the coupling between the localized system and the same field (i.e., the losses). \\
A noteworthy detail is that the Hamiltonian $\hat{H}$ contains the \textit{localized} system terms only, i.e., it does not depend on field terms. As an example, $\hat{H}$ might be one of the Hamiltonian operators in Eqs.~(\ref{eq: SYSTEM Hamiltonians}), i.e., it may contain  the Hamiltonian terms describing the free evolution of the localized systems and the time-dependent interaction Hamiltonian between them. The input parameters of the model would be, in this case, the detunings $\Delta_s$, $\Delta_c$, and $\Delta_{s,c}$, as well as the qubit-cavity interaction strength, $g$. On the other hand, the system-field interaction Hamiltonian are embedded in the set of differential Eqs.~(\ref{eq: Master equation for field in Fock state}), in which the single-photon (time-reversed) wavepackets $\xi_p(t)$ ($p = s,c$) appear together with additional coefficients $m,n$ due to the action of the field operators $\hat{b}, \hat{b}^\dagger$ on the Fock states of the field \cite{Baragiola:2014} (see equations below). Also, the coefficients $\gamma_i$ in Eqs.~(\ref{eq: Interaction Hamiltonians, system - fields, appendix}) define the coupling operators, $\mathbf{L}$. \\

A characteristic of the SLH framework is the possibility to describe a system with \textit{multiple} input-output channels. In such a case, the scattering and coupling operators are truly represented via matrices and vectors (of operators): each component of these matrices (or vectors) refers to a single input-output channel. This can be seen, for example, in Eq.~(\ref{eq: SLH triple of the switch}). Besides, when dealing with \textit{multiple} open quantum systems, it is possible to \textit{compose} the triples of each system \cite{GoughJames_2008, Gough_James_2009}, thus obtaining an overall triple that fully describes the entire ensemble network of localized systems coupled by itinerant field (the network). 

The SLH triple of the whole network completely defines a master equation for the state of the localized system, $\hat{\rho}(t)$. Therefore, in order to determine and solve the appropriate set of coupled equations, the first step is to  find the correct triple that describes the network, $\mathbf{G}$, i.e., 
\begin{equation}
\begin{split}
\frac{d}{dt}\hat{\rho}(t) = &\mathcal{L}[\hat{\rho}(t)] = f(\mathbf{S}, \mathbf{L}, \hat{H}, \hat{\rho}(t), \xi(t)) \\
&\iff \mathbf{G} = (\mathbf{S}, \mathbf{L}, \hat{H}) \, .
\end{split}
\end{equation}
Here, $\xi(t)$ is the time-dependent (and time-reversed) pulse wavepacket, and $\mathcal{L}[\bullet]$ indicates the Liouville superoperator, whose general expression is:
\begin{equation}
    \label{eq: Lindbladian, generic}
    \mathcal{L}[\hat{L}]\hat{\rho} \equiv \sum_{i=0}^{N_\text{c}}\hat{L}_i\hat{\rho}\hat{L}_i^\dagger -\frac{1}{2}\bigl(\hat{L}_i^\dagger\hat{L}_i \hat{\rho} + \hat{\rho} \hat{L}_i^\dagger\hat{L}_i \bigr) \, ,
\end{equation}
in which $N_\text{c}$ is the number of input-output channels of the network. The function  $f(\bullet)$ represents a function of the triple $\mathbf{G}$, but also depending on  $\hat{\rho}(t)$ and $\xi(t)$. Ultimately, this function is uniquely determined by $\mathbf{G}$. \\
Now, let us consider an environment whose initial state is the Fock state $\ket{N}$ (with $N\neq 0$): therefore, after the interaction with the localized system, the output field state will be considerably changed, thus compromising the Markov hypothesis. Hence, an extension to the formalism has been provided \cite{Baragiola:2014, Baragiola_article}: the triple $\mathbf{G}$ corresponds now to a \textit{set} of coupled differential equations for the \textit{generalized density operator} $\hat{\rho}_{\mathbf{m}, \mathbf{n}}(t)$, defined as:
\begin{equation}
\label{eq: gen density operator, appendix slh framework}
\hat{\rho}_{\mathbf{m}, \mathbf{n}}(t) \equiv \text{Tr}_{\text{field}}[\hat{U}(t)(\hat{\rho}_{\text{sys}}(0)\otimes\hat{\rho}_{\mathbf{m}, \mathbf{n}}^{\text{field}})\hat{U}^\dagger(t)] \, ,
\end{equation}
in which $\mathbf{m}, \mathbf{n} = m_1m_2...m_{N_\text{c}},n_1n_2...n_{N_\text{c}}$: each $m_i$, $n_i$ represents the number of photons present the $i$-th field mode (i.e., the $i$-th input-output channel), being $N_\text{c}$ the total number of modes (i.e., the number of input-output channels). We notice that $m_i$ can be different from $n_i$, since we can consider field operators $\ket{m_1m_2...m_{N_\text{c}}}\bra{n_1n_2...n_{N_\text{c}}}$, which are non-diagonal in the Fock state basis. \\
As an illustrative example, if we assume to deal with a single input-output channel, we can simplify to $\mathbf{m}, \mathbf{n} = m,n$. In this case, the set of coupled differential equations to be solved reduces to
\begin{equation}
\begin{aligned}
        \label{eq: Master equation for field in Fock state}
        \frac{d}{dt}&\hat{\rho}_{m,n}(t) = -\frac{i}{\hbar}\big[\hat{H},\hat{\rho}_{m,n}(t)\big] +\mathcal{L}[\hat{L}]\hat{\rho}_{m,n}(t) \\
        &+ \sqrt{m}\xi(t)\big[\hat{S}\hat{\rho}_{m-1,n}(t), \hat{L}^{\dagger}\big] \\
        &+\sqrt{n}\xi^{\ast}(t)\big[\hat{L}, \hat{\rho}_{m,n-1}(t) \hat{S}^{\dagger}\big]   \\
        & + \sqrt{mn}|\xi(t)|^2\big(\hat{S}\hat{\rho}_{m-1,n-1}(t)\hat{S}^{\dagger}-\hat{\rho}_{m-1,n-1}(t)\big) \, .
\end{aligned}
\end{equation}

\subsection{SLH triple of the single-photon switch}
We now specify the SLH triple used to describe the single-photon switching device discussed in the present work, which can be assembled by considering three input-output channels. Essentially, each channel is associated with a trivial scattering operator, $\hat{S}_i = \hat{\mathds{I}}$ (here, $\hat{\mathds{I}}$ is the identity operator on the given Hilbert space dimension), and a coupling operator $\hat{L}_i = \sqrt{\gamma_i}\hat{a}_i$ (with $\hat{a}_i = \hat{a}_1, \hat{a}_2$ or $\hat{\sigma}_-$). We hereby report the SLH triples of the three-channel networks defined for ``Model A'' and ``Model B'', respectively:
\begin{equation}
    \label{eq: SLH triple of the switch}
    \begin{split}
    \mathbf{G}^A_{\text{switch}} &= \Biggl( 
    \begin{bmatrix}
        \hat{\mathds{I}} & \hat{\mathds{O}} & \hat{\mathds{O}} \\
        \hat{\mathds{O}} & \hat{\mathds{I}} & \hat{\mathds{O}} \\
        \hat{\mathds{O}} & \hat{\mathds{O}} & \hat{\mathds{I}} \\
    \end{bmatrix}, \quad  
     \begin{bmatrix}
        \sqrt{\gamma_1}\hat{a}_1 \\
        \sqrt{\gamma_2}\hat{a}_1 \\
        \sqrt{\gamma_3}\hat{\sigma}_- \\
    \end{bmatrix}, \quad \hat{H}_{\text{sys}}^A
    \Biggr)  \\
    \mathbf{G}^B_{\text{switch}} &= \Biggl( 
    \begin{bmatrix}
        \hat{\mathds{I}} & \hat{\mathds{O}} & \hat{\mathds{O}} \\
        \hat{\mathds{O}} & \hat{\mathds{I}} & \hat{\mathds{O}} \\
        \hat{\mathds{O}} & \hat{\mathds{O}} & \hat{\mathds{I}} \\
    \end{bmatrix}, \quad  
     \begin{bmatrix}
        \sqrt{\gamma_1}\hat{a}_1 \\
        \sqrt{\gamma_2}\hat{a}_2 \\
        \sqrt{\gamma_3}\hat{\sigma}_- \\
    \end{bmatrix}, \quad \hat{H}_{\text{sys}}^B
    \Biggr) \, ,
    \end{split}
\end{equation}
in which $\hat{H}_{\text{sys}}^X$ (with $X=A,B$) represents the system Hamiltonian according to one of the models in  Eq.~(\ref{eq: SYSTEM Hamiltonians}), and $\hat{\mathds{O}}$ is a matrix of zeroes. \\
For the results shown in this work, the input field modes of channel 1 (the signal input) and channel 3 (the control input) are initialized to the Fock state $\ket{1}$. On the other hand, the input field mode of channel 2 (the output of the switch) is set to the vacuum state $\ket{0}$. 

\section{SLH evolution equations \\ for the single-photon switch}\label{appendix: slh equations}

\subsection{Evolution equations for the density operators}
As it has been shown in Refs.~\cite{Baragiola_article, Baragiola:2014}, all the elements discussed above can be employed to derive a set of coupled differential equations allowing to describe the temporal evolution of the generalized density operators $\hat{\rho}_{\mathbf{m}, \mathbf{n}}(t)$ defined in Eq.~(\ref{eq: gen density operator, appendix slh framework}). In our specific case, $\hat{\rho}_{\mathbf{m}, \mathbf{n}}^{\text{field}} = \ket{m_1m_2m_3}\bra{n_1n_2n_3}$ and $m_1,m_3,n_1,n_3$ can be either 0 or 1, while $m_2,n_2 = 0$. Here, $\hat{U}(t)$ is the unitary evolution operator of the whole system (i.e., localized systems and environment). \\
The general expression of the differential equations for the $(\mathbf{m},\mathbf{n})$-th generalized density operator is thus given as
\begin{equation}
\label{eq: SLH coupled differential equations of the switch}
    \begin{split}
        &\frac{d}{dt}\hat{\rho}_{\mathbf{m},\mathbf{n}}(t) = -\frac{i}{\hbar}\big[\hat{H}_{\text{sys}},\hat{\rho}_{\mathbf{m},\mathbf{n}}(t)\big] +\sum_{i=1}^{3}\mathcal{L}[\hat{L}_i]\hat{\rho}_{\mathbf{m},\mathbf{n}}(t)  \\
        &+ \sum_{i,j=1}^{3}\sqrt{m_j}\xi_j(t)\big[\hat{S}_{ij}\hat{\rho}_{m_j - 1,\mathbf{n}}(t), \hat{L}_i^{\dagger}\big]  \\
        &+\sum_{i,j=1}^{3}\sqrt{n_j}\xi^{\ast}_j(t)\big[\hat{L}_i, \hat{\rho}_{\mathbf{m},n_j - 1}(t) \hat{S}_{ij}^{\dagger}\big]   \\
        &+ \sum_{i,j=1}^{3}\sqrt{m_i n_j}\xi_i^{\ast}(t)\xi_j(t) \times   \\
       &\times \Big[\sum_{k=1}^{3}\hat{S}_{kj}\hat{\rho}_{m_i - 1 , n_j - 1}(t)\hat{S}_{ki}^{\dagger}
        - \delta_{ij}\hat{\rho}_{m_i - 1 , n_j - 1}(t)\Big]  \, ,
    \end{split}
\end{equation}
in which $\hat{\rho}_{m_j - 1,\mathbf{n}}$ denotes the sum of three contributions: $\hat{\rho}_{(m_1 - 1)m_2m_3,\mathbf{n}}$, $\hat{\rho}_{m_1(m_2 - 1)m_3,\mathbf{n}}$, and $\hat{\rho}_{m_1m_2(m_3 - 1),\mathbf{n}}$, leaving $\mathbf{n}=n_1n_2n_3$ untouched. \\
Each $m_j, n_j$ refers to the labels of the density operator $\hat{\rho}_{\mathbf{m}, \mathbf{n}}(t)$ associated with each specific evolution equation. As an example, in the case of the equation for $\hat{\rho}_{101,100}(t)$ we have $m_1 = 1$, $m_2 = 0$, $m_3 = 1$ and $n_1 = 1$, $n_2 = 0$, $n_3 = 0$; moreover, the terms multiplied by $\sqrt{m_j}$ vanish if $m_j=0 \text{ }\forall j$, and the same holds true also for $\sqrt{n_j}$. \\
Besides, we remark that the operators $\hat{S}_{ij}$, $\hat{L}_i$ and $\hat{H}_{\text{sys}}$ are the ones that show up in the single-photon switch triple, see Eq.~(\ref{eq: SLH triple of the switch}). Also, $\xi_i(t) = \xi_s(t)$ (i.e., the signal single-photon wavepacket) if $i=1,2$, while $\xi_i(t) = \xi_c(t)$ (the control single-photon wavepacket) if $i=3$, as defined in Eq.~(\ref{eq: time-reversed wavepacket, appendix}). \\
Each of the equations defined in (\ref{eq: SLH coupled differential equations of the switch}) can be solved numerically by starting from the lowest-level equation, which is the one relative to the generalized density operator $\hat{\rho}_{000,000}(t)$, and then proceeding hierarchically. In fact, once the solution for $\hat{\rho}_{000,000}(t)$ is found, the higher-level coupled differential equations can be solved by exploiting the property $\hat{\rho}^\dagger_{\mathbf{n}, \mathbf{m}}(t) = \hat{\rho}_{\mathbf{m}, \mathbf{n}}(t)$. Finally, the result is the generalized density matrix $\hat{\rho}_{101,101}(t)$ at each evolution time step, $t$, which describes the state of the whole localized system (i.e., Cavity 1 + Cavity 2 + qubit in model B), given that the states of the input-output channel modes 1,2, and 3 have been initialized as $\ket{1}, \ket{0}$ and $\ket{1}$, respectively.\\
Ultimately, after obtaining the numerical solution for $\hat{\rho}_{101,101}(t)$, it is possible to calculate the expectation values of any generic system operator $\hat{X}(t)$ at time $t$, defined as $\langle \hat{X}(t) \rangle = \text{Tr}_{\text{sys}}[\hat{\rho}_{101,101}(t)\hat{X}(0)]$. 

\subsection{The photon flux}

One of the main quantities used in this work to show the functionalities of the single-photon switch involves calculatin the averaged photon flux through a specific output channel. Here we follow the procedure outlined in Ref.~\cite{Baragiola_article}. Indeed, the photon flux through a given output is a function of the generalized density operators, $\hat{\rho}_{\mathbf{m}, \mathbf{n}}(t)$, the triple of operators $\mathbf{S}, \mathbf{L}$, and $\hat{H}_{\text{sys}}$ defined in  Eq.~(\ref{eq: SLH triple of the switch}), and the given single-photon time reversed wavepackets $\xi_p(t)$. \\
We first define the average photon flux at the output of channel $i$ as $\phi_i(t)$. This is calculated as \cite{Baragiola_article}
\begin{equation}
        \label{eq: mean photon flux MULTIMODE}
        \begin{split}
        &\phi_i(t) = \mathds{E}_{\mathbf{m},\mathbf{n}}[\hat{L}_i^{\dagger}(t)\hat{L}_i(t)] \\
        &+\sum_{k=1}^{3}\sqrt{n_k}\xi_k(t)\mathds{E}_{\mathbf{m},n_k - 1}[\hat{L}_i^{\dagger}(t)\hat{S}_{ik}(t)] \\
        &+ \sum_{k=1}^{3}\sqrt{m_k}\xi_k^{\ast}(t)\mathds{E}_{m_k - 1,\mathbf{n}}[\hat{S}_{ik}^{\dagger}(t)\hat{L}_i(t)]\\
    &+ \sum_{k,l=1}^{3} \sqrt{m_k n_l}\xi_k^{\ast}(t) \xi_l(t) \mathds{E}_{m_k - 1,n_l - 1}[\hat{S}_{ik}^{\dagger}(t)\hat{S}_{il}(t)] \, ,
        \end{split}
    \end{equation}
in which $\mathds{E}_{\mathbf{m},\mathbf{n}}[\mathcal{\hat{O}}]$ indicates the \textit{asymmetric} expectation of the overall system (i.e., localized system and environment) operator $\mathcal{\hat{O}}$ \cite{Baragiola:2014}
\begin{equation}
        \label{eq: asymmetric expectation}
        \mathds{E}_{\mathbf{m},\mathbf{n}}[\mathcal{\hat{O}}] = \text{Tr}_{\text{sys}}\big[\hat{\rho}^\dagger_{\mathbf{m}, \mathbf{n}}(t)\mathcal{\hat{O}}\big] \, .
\end{equation}
Finally, the integrated photon flux up to time $t$ is defined as
\begin{equation}
    \label{eq: integrated photon flux APPENDIX}
    \Phi_i(t) \equiv \int_{t_0}^{t}\phi_i(t')dt' \, ,
\end{equation}
assuming $t_0$ as the integration starting time. \\
In summary, once the operators $\hat{\rho}_{\mathbf{m}, \mathbf{n}}(t)$ have been calculated by solving the relative set of coupled differential equations, we can first proceed to determine the averaged photon flux, $\phi_i(t)$, for each channel and, then, integrate it to finally obtain $\Phi_i(t)$. 
In case of time-independent parameters $\gamma_i$, the overall integrated flux through all of the input-output channels is calculated as $\Phi(t) = \sum_i \Phi_i(t) \rightarrow 2$ for $t\rightarrow +\infty$, since only two photons are present in the network in our case. 
As a final consideration, we remark that all the numerical integrations reported in this work have been carried out via the standard Runge-Kutta IV algorithm in an originally written Python code.

\section{Maximizing the photon flux \\ for time-dependent output rates}
\label{appendix: numerical opt, time-dependent}
As described in the main text, we have numerically found that the single-photon switch performance can be improved by extending the model to a time-dependent description. In order to do so, it is useful to introduce time-dependent coupling rates to the output channels, i.e., $\gamma_i(t)$. A similar attempt has been already proposed in Ref.~\cite{xu2016single}, although with a different approach. Here we assume that a linear coupling between the localized systems and the field modes still holds, although its intensity is allowed to vary with time, as it has already been shown experimentally for microwave superconducting circuits Ref.~\cite{yin_catch_2013}. We also notice that such an approximation is theoretically supported within the SLH framework, for which we can still maintain the same formalism by considering a time-dependent SLH triple, $(\mathbf{S}(t), \mathbf{L}(t), \hat{H}(t))$. For a more detailed account of this case, we refer to  Refs.~\cite{nurdin_perfect_2016,gough_generating_2015}, as well as Refs.~\cite{gough_quantum_2012, gough_quantum_2011, kiilerich_input-output_2019, kiilerich_quantum_2020}. \\
 Mathematically speaking, the time dependence can be modelled by considering coupling parameters defined as $\gamma_i(t) = \gamma_i \Theta(t-t_0)$ or $\gamma_i(t)= \gamma_i\Theta(t-t_0)\Theta(t_1-t)$, in which $\gamma_i$ are time-independent parameters parameters. We hereby conjecture that these parameters can be chosen as the values previously optimized with a time-independent theory. Hence, the modulated system-field interactions can be formulated analogously to Eqs.~(\ref{eq: Interaction Hamiltonians, system - fields, appendix}), i.e., formally becoming
\begin{equation}
    \label{eq: Interaction Hamiltonians, system - fields, TIME-DEPENDENT, appendix}
    \begin{split}
    &\hat{H}^{(\text{cavity 1, } s)}_{\text{int}} = i\hbar\sqrt{\gamma_1(t)}[\hat{a}_1\hat{b}_s^\dagger(t) - \hat{a}_1^\dagger \hat{b}_s(t)] \\
    &\hat{H}^{(\text{cavity 1, } o)}_{\text{int}} = i\hbar\sqrt{\gamma_2(t)}[\hat{a}_1\hat{b}_o^\dagger(t) - \hat{a}_1^\dagger \hat{b}_o(t)] \\
    &\hat{H}^{(\text{cavity 2, } o)}_{\text{int}} = i\hbar\sqrt{\gamma_2(t)}[\hat{a}_2\hat{b}_o^\dagger(t) - \hat{a}_2^\dagger \hat{b}_o(t)] \\
    &\hat{H}^{(\text{qubit, } c)}_{\text{int}} = i\hbar\sqrt{\gamma_3(t)}[\hat{\sigma}_-\hat{b}_c^\dagger(t) - \hat{\sigma}_+ \hat{b}_c(t)] \, .
    \end{split}
\end{equation}
Here, we provide the details regarding the different network configurations that we tested, in order to increase the efficiency of the device by employing such time-dependent parameters. \\
For each configuration, the steady state photon flux difference $\Phi_2^{(s,c)} - \Phi_2^{(c)}$ for the output channel has been maximized, by focusing on the simplest model A. Only in the last configuration, we compared results obtained when considering a network described by model A to those obtained for the full network described by model B. From the comparison, it becomes evident that a filtering cavity is necessary for the realization of a realistic and performing single-photon switching. 

\begin{figure}[t]
    \centering
    \includegraphics[width=0.9\linewidth]{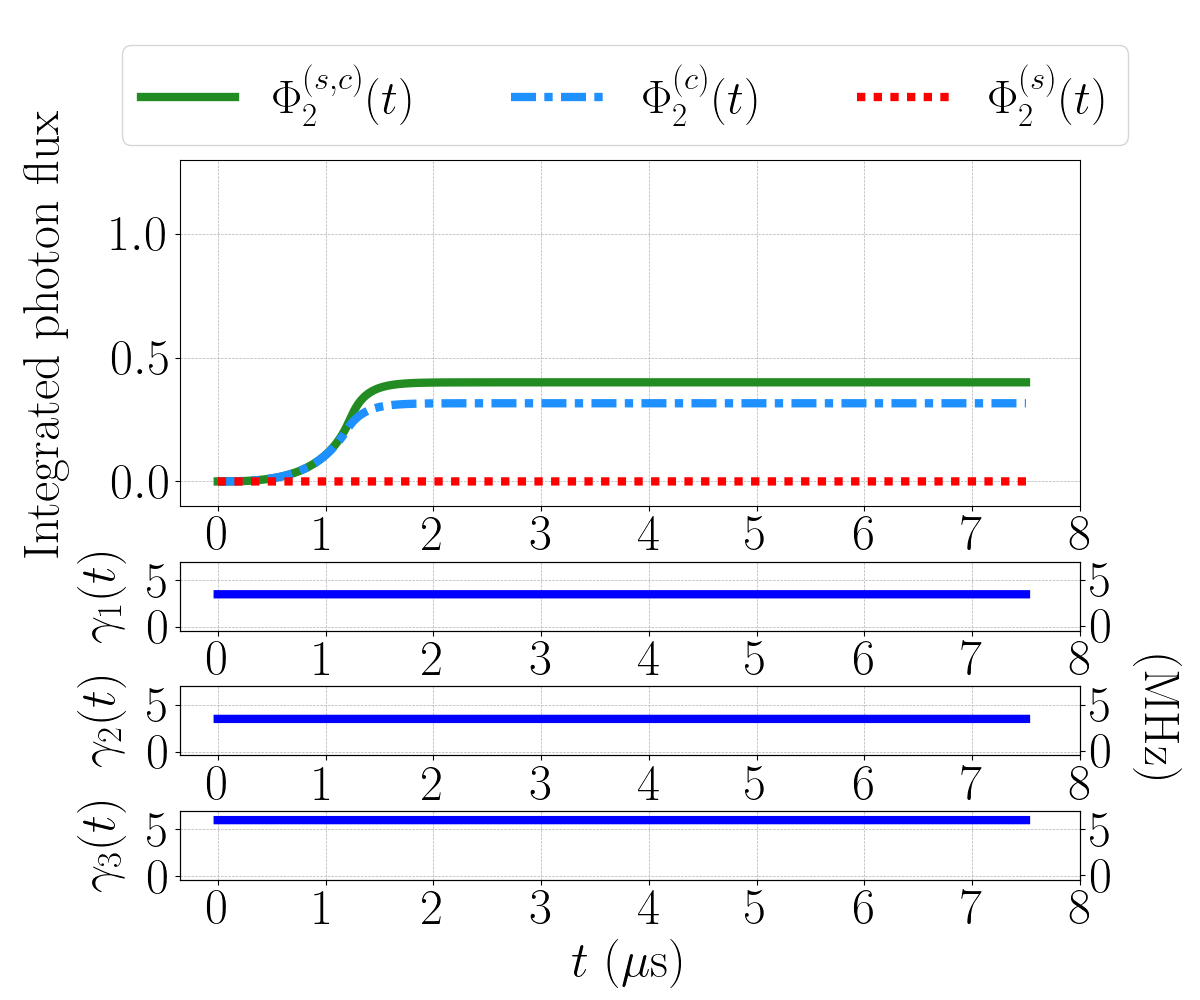}
    \caption{Time-independent coupling rates $\gamma_i$, in model A. The integrated photon fluxes $\Phi_2^{(s,c)}$, $\Phi_2^{(c)}$, and $\Phi_2^{(s)}$ are plotted versus time; for completeness, the (constant) parameters $\gamma_i(t)$ are also shown in the bottom panels. In this case, the long time behavior of the flux difference gives $\Phi_2^{(s,c)} - \Phi_2^{(c)}= 0.085$.}
    \label{fig: FLUX AND GAMMA 1}
\end{figure}

\begin{figure}
    \centering
    \includegraphics[width=0.9\linewidth]{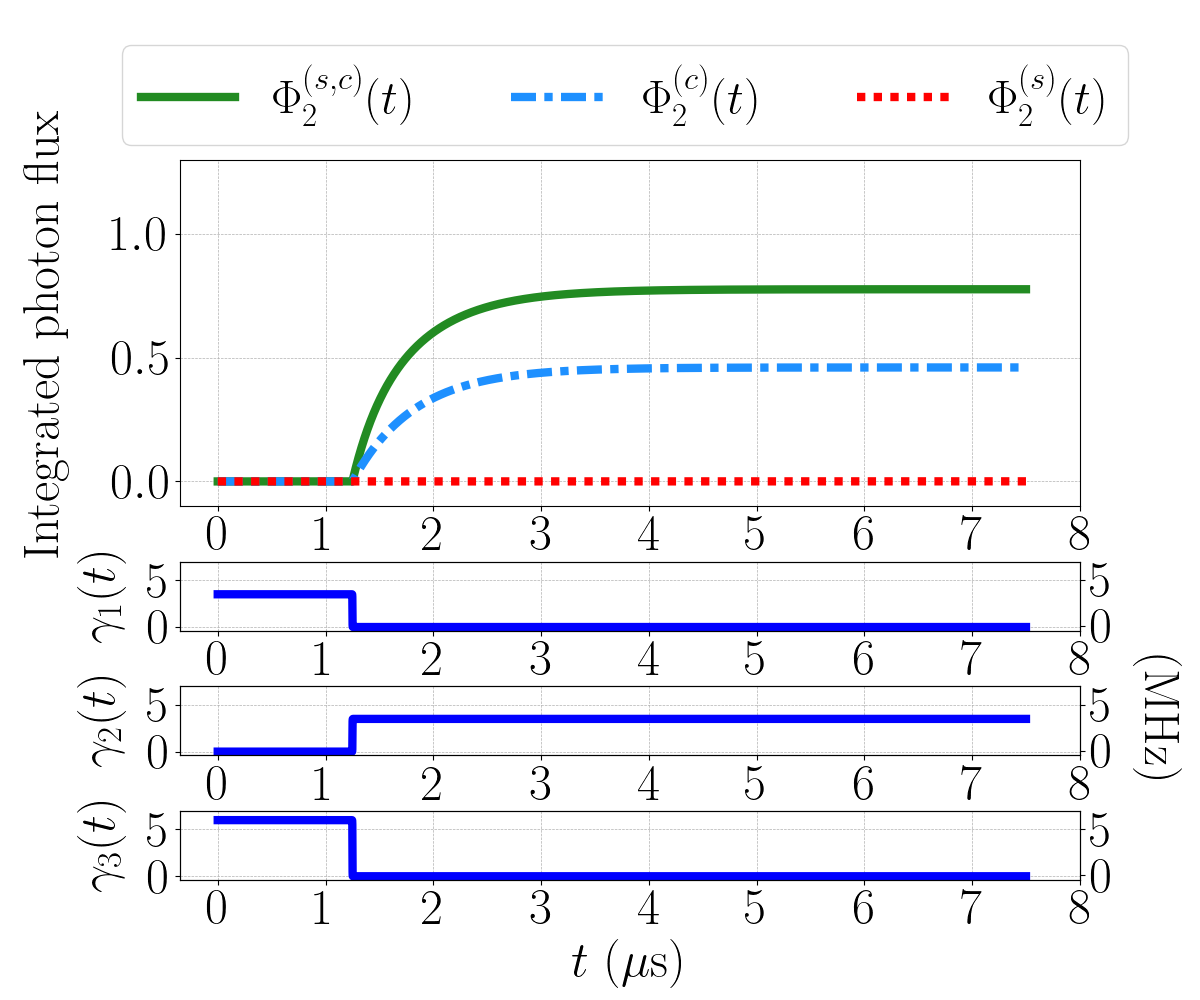}
    \caption{Steady state fluxes and time-dependent coupling rates from Eqs.~(\ref{eq: time-dep gammas, STEP 1, appendix}), used to solve the dynamical equations of model A. The calculated flux difference in steady state is $\Phi_2^{(s,c)} - \Phi_2^{(c)}= 0.316$.}
    \label{fig: FLUX AND GAMMA 2}
\end{figure}

Let us start from the situation with time-independent coupling parameters, reported in Fig.~\ref{fig: FLUX AND GAMMA 1}: in this case, the calculated flux difference was about $\Phi_2^{(s,c)} - \Phi_2^{(c)}= 0.085$ in steady state (i.e., long integration times). In view of maximizing this quantity, we first choose the following piecewise constant functions of time as a first attempt: 
\begin{equation}
    \label{eq: time-dep gammas, STEP 1, appendix}
    \begin{split}
    \gamma_1(t) &= \gamma_1\Theta(t_a^{(s)}-t) \\
    \gamma_2(t) &= \gamma_2\Theta(t - t_a^{(s)}) \\
    \gamma_3(t) &= \gamma_3\Theta(t_a^{(s)} - t) \, .
    \end{split}
\end{equation}
According to this choice, $\gamma_1(t)$ and $\gamma_3(t)$ are switched off when $\gamma_2(t)$ is switched on, after the signal photon's arrival. We notice again that the parameter values in Eqs.~(\ref{eq: time-dep gammas, STEP 1, appendix}), as well as in the following,  are assumed equal to the time-independent case. Experimentally, the crossover from $\gamma_i(t)\approx 0$ and $\gamma_i(t)\approx \gamma_i$ can be performed within a $\sim 10 \text{ ns}$ timescale \cite{yin_catch_2013, pechal2016superconducting}. 
In this case, the difference between the fluxes converges to the maximal value $\Phi_2^{(s,c)} - \Phi_2^{(c)}\rightarrow0.316$. This result can be appreciated by looking at Fig.~\ref{fig: FLUX AND GAMMA 2}, and by comparing it to the time-independent case shown in Fig.~\ref{fig: FLUX AND GAMMA 1}.

\begin{figure}
    \centering
    \includegraphics[width=0.9\linewidth]{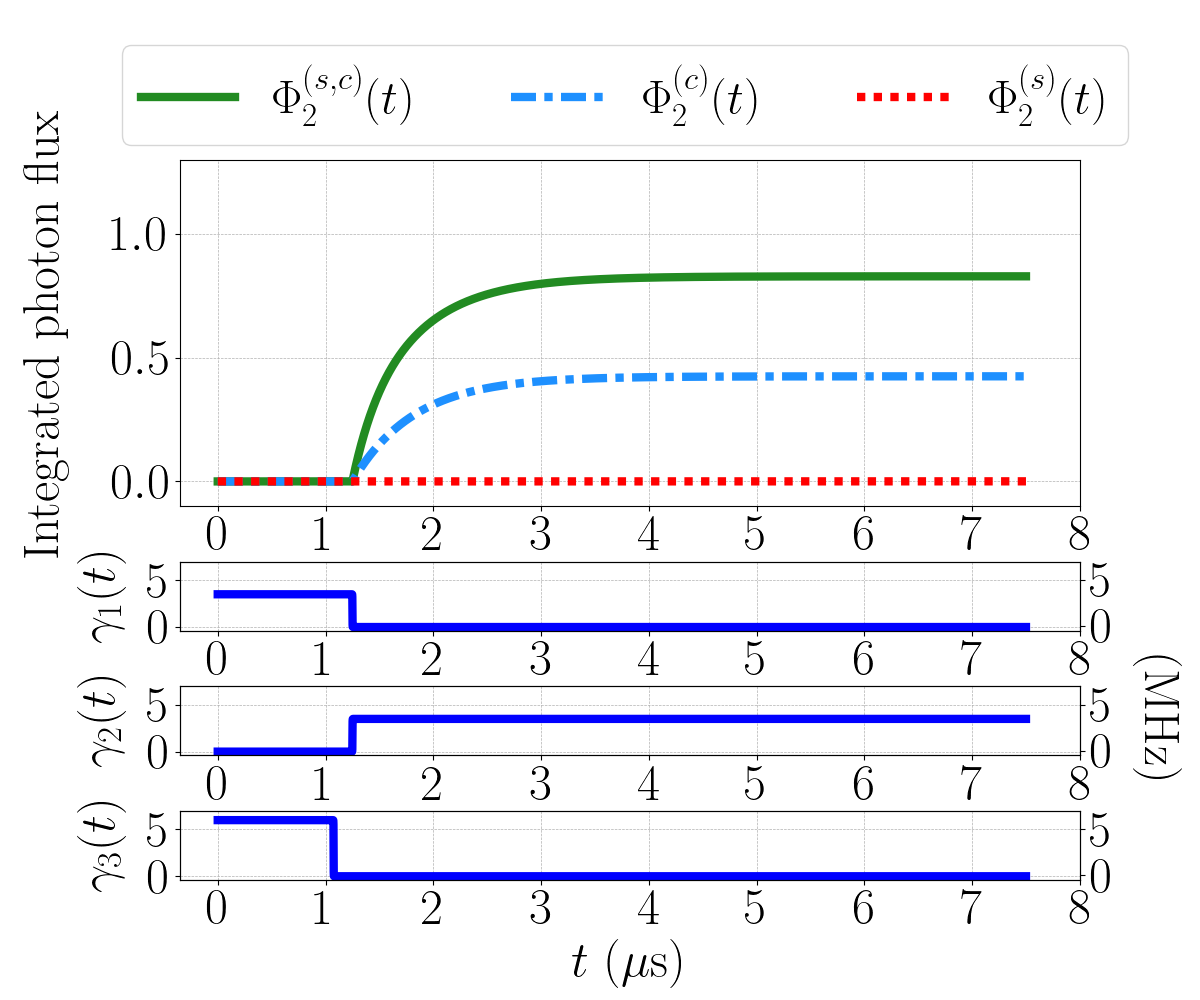}
    \caption{Photon fluxes in steady state  and time-dependent coupling rates from  Eqs.~(\ref{eq: time-dep gammas, STEP 2, appendix}), used to solve the dynamical equations of model A; the calculated flux difference for this optimized case results as $\Phi_2^{(s,c)} - \Phi_2^{(c)}= 0.404$.}
    \label{fig: FLUX AND GAMMA 3}
\end{figure}

As a next step, we have tuned the arrival time  of the single control photon, $t_a^{(c)}$, by anticipating it with respect to the signal photon arrival, fixed as $\omega_0 t_a^{(s)} = 5$ (i.e., $t_a^{(s)} = 1.25$~$\mu$s). In addition, $\gamma_3(t)$ is turned off when $t = t_a^{(c)}$: 
\begin{equation}
    \label{eq: time-dep gammas, STEP 2, appendix}
    \begin{split}
    \gamma_1(t) &= \gamma_1\Theta(t_a^{(s)}-t) \\
    \gamma_2(t) &= \gamma_2\Theta(t - t_a^{(s)}) \\
    \gamma_3(t) &= \gamma_3\Theta(t_a^{(c)} - t) \, .
    \end{split}
\end{equation}
Here, we numerically find that the flux difference is maximized to 0.404 for $\omega_0t_a^{(c)} = \omega_0t_0 = 4.3$ (which corresponds to $1.075$~$\mu$s), as reported in Fig.~\ref{fig: FLUX AND GAMMA 3}. 

After these attempts, we notice that letting the control photon arrive earlier may yield detrimental effects. In fact, once entered in the JC system, the control photon could escape from the input channel (i.e., the one from which the signal photon is injected) before the arrival of the signal photon itself. This is possible, in principle, since the coupling coefficient $\gamma_1(t)$ is nonzero for times smaller than $t_a^{(s)}$. Therefore, we have introduced a different time shape for $\gamma_1(t)$, in order to minimize this undesired process:
\begin{equation}
    \label{eq: gamma1 final, appendix}
    \gamma_1(t) = \gamma_1\Theta(t-t_0)\Theta(t_a^{(s)} - t) \, .
\end{equation}
This temporal dependence corresponds to the ultimate configuration already analyzed in the manuscript, i.e., the one related to Eqs.~(\ref{eq: time-dep gammas, MAIN}). Here, $\gamma_1(t)$ is a square function that mimics a fast on/off coupling between times $t_0$ (the turning-on time) and $t_a^{(s)}$ (i.e., when $\gamma_1(t)$ turns off). This way the input channel is opened after a certain time $t_0$, and closed only when the signal photon arrives, at $t = t_a^{(s)}$. This prevents the control photon from escaping through this channel long before the signal photon arrival.
Hence, by exploiting this time-dependent coupling factor we could also optimize the parameter $t_0$ in Eq.~(\ref{eq: gamma1 final, appendix}). Here, $\omega_0 t_a^{(s)} = 5$ ($t_a^{(s)} = 1.25$~$\mu$s) is fixed, while $t_0$ is varied (by keeping $t_0 < t_a^{(s)}$). The optimal value is numerically found as $\omega_0 t_0 = 3.7$ (i.e., $t_0 = 0.925$~$\mu$s), for which we obtain the best numerical result (in the absence of any filtering cavity) for the difference in photon fluxes: $\Phi_2^{(s,c)} - \Phi_2^{(c)}\rightarrow0.489$ in steady state. 
\begin{figure}[h]
    \centering
    \includegraphics[width=0.9\linewidth]{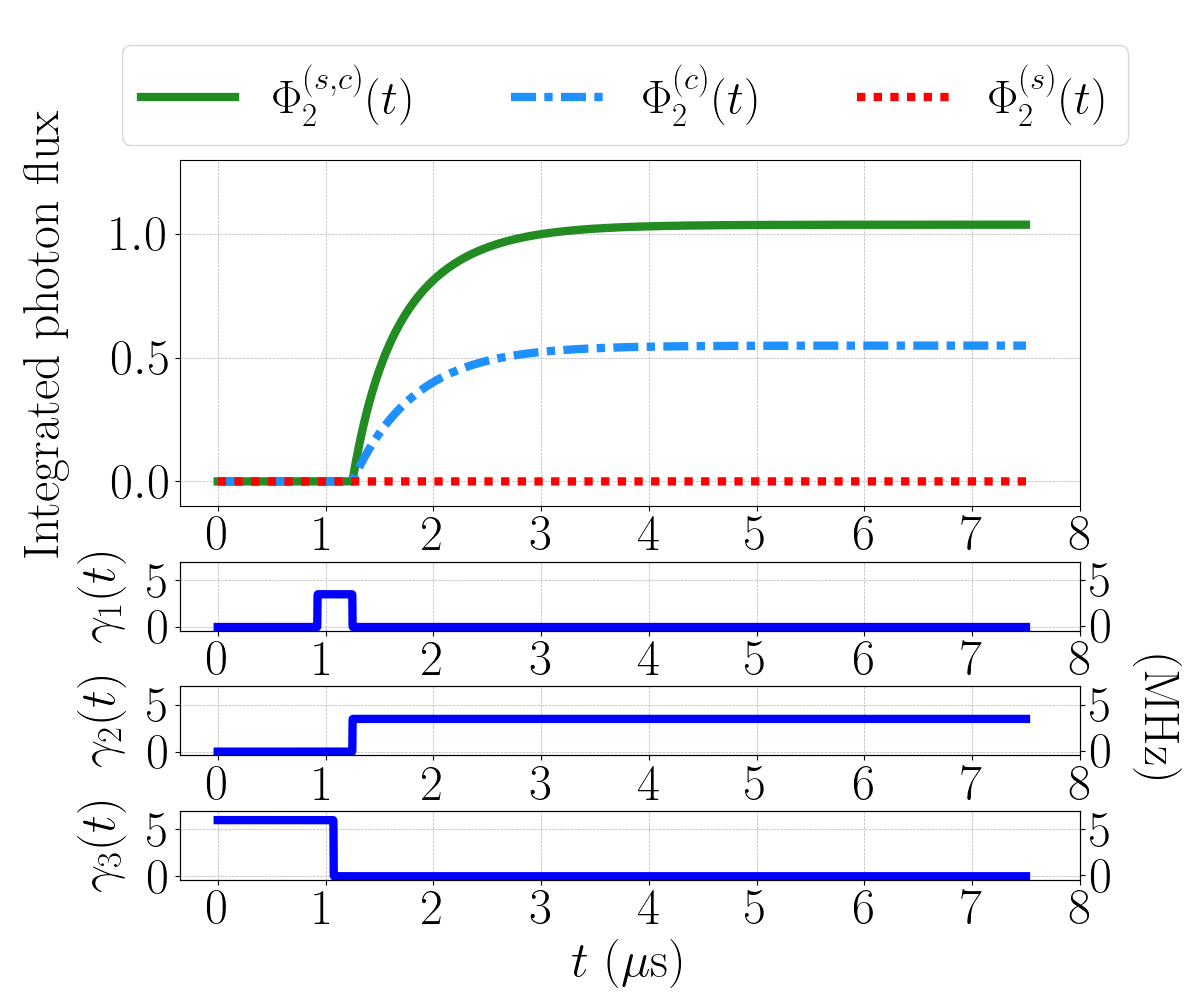}
    \caption{Steady state photon fluxes and time-dependent coupling rates from Eqs.~(\ref{eq: time-dep gammas, MAIN}), solving dynamical equations for model A. This is the best result obtained in this work in terms of the flux difference, i.e. $\Phi_2^{(s,c)} - \Phi_2^{(c)}= 0.489$ in steady state.}
    \label{fig: FLUX AND GAMMA 4}
\end{figure}

Finally, the whole device performance is optimized by maximizing the signal photon content in the output channel. In this case, we take into account model B by adding the second  cavity (Cavity 2, or filtering cavity) weakly coupled to the first one, and tune its frequency into resonance with the signal pulse carrier frequency, as schematically shown in Fig.~\ref{fig:model}(b). We then numerically scan the coupling rate between the two resonators, $J$, to maximize the output flux difference. \\
From these simulations, we have found that, for high values of $J$, $\Phi_2^{(s,c)} - \Phi_2^{(c)}$ reduces since a stronger coupling between the cavities modifies the JC spectrum of the first cavity coupled to the qubit. Under such conditions the operating principle of the switch tends to break down. 
However, if the filtering cavity is only weakly coupled to the first cavity, the flux difference goes back to values that are comparable to the optimal results found before. In particular, we have numerically found that $J = 1\text{ MHz}$ is an optimal value, leading to $\Phi_2^{(s,c)} - \Phi_2^{(c)}\rightarrow0.429$: this is the best case of model B, which has been reported in the main text, in Fig.~\ref{fig: switch process}.

\end{document}